\documentclass[draft]{agujournal2019}
\usepackage[T1]{fontenc}
\usepackage{fix-cm}
\usepackage{url} 
\usepackage{lineno}
\usepackage[finalnew]{trackchanges} 
\usepackage{soul}

\usepackage{amsmath,amsfonts,amssymb,amsthm}
\usepackage{graphicx}
\graphicspath{{./}}

\draftfalse

\journalname{JGR: Space Physics}

\newcommand{\grl}{    {Geophys. Res. Lett.}}
\newcommand{\jgr}{    {J. Geophys. Res.}}

\newcommand{\wpi}{\omega_{pi}}
\newcommand{\wpe}{\omega_{pe}}

\newcommand{\Gn}{\Gamma_n}
\newcommand{\Gnp}{\Gamma_n'}

\begin{document}

%
%


\title{Ion pickup and velocity space thermalization at outer planet moons}

%
%




\authors{Xin An\affil{1,2,3}, Miranda Chang\affil{3}, Hao Cao\affil{3}, Vassilis Angelopoulos\affil{3}, Anton Artemyev\affil{4,3}}


\affiliation{1}{Department of Space Science, University of Alabama in Huntsville, Huntsville, AL, 35805, USA}
\affiliation{2}{Center for Space Plasma and Aeronomic Research (CSPAR), The University of Alabama in Huntsville, Huntsville, AL 35805, USA}
\affiliation{3}{Department of Earth, Planetary, and Space Sciences, University of California, Los Angeles, CA, USA}
\affiliation{4}{Department of Physics, University of Texas at Arlington, Arlington, TX, USA}




\correspondingauthor{Xin An}{phyax1@gmail.com}



\begin{keypoints}
\item Transverse and compressional magnetic perturbations convert bulk kinetic to thermal energy, completing ion pickup within a few gyroperiods.
\item Transverse perturbations are EMIC waves while compressional perturbations are ion Bernstein and mirror-mode waves.
\item We identify resonances and velocity-space structures responsible for wave growth and damping via field-particle correlation analysis.
\end{keypoints}

%
%

%
%


\begin{abstract}
Ion pickup at the outer planets' active moons is a fundamental plasma process in which newly ionized particles from moon exospheres interact with the ambient corotating plasma and are accelerated to match the background flow. Spacecraft observations have revealed intense electromagnetic wave activity commonly attributed to this pickup process. Here we investigate ion pickup using hybrid-kinetic simulations in which ions are treated kinetically while electrons are modeled as a massless fluid. In the moon's rest frame, ambient ions initially stream perpendicular to the background magnetic field at the corotation velocity, creating a nongyrotropic velocity distribution with two ion populations clustered at opposite gyrophases. Within a few ion gyroperiods, this configuration simultaneously excites transverse magnetic perturbations associated with electromagnetic ion cyclotron waves and compressional perturbations associated with mirror-mode and ion Bernstein waves, reaching amplitudes of several percent of the background field strength. Using field-particle correlation analysis, we quantify the energy transfer between waves and particles and demonstrate how these perturbations scatter ions in velocity space, efficiently incorporating newly created ions into the background plasma and leading to isotropization in both gyrophase and pitch angle. These results provide a kinetic framework for understanding pickup-driven wave-particle interactions and offer guidance for interpreting in situ measurements at active moons throughout the outer solar system.



\end{abstract}

\section*{Plain Language Summary}
Several moons orbiting Jupiter and Saturn have thin atmospheres that constantly leak particles into space. When these particles become electrically charged (ionized), they interact with the surrounding plasma environment flowing past the moon, creating intense electromagnetic waves that spacecraft have detected. We used computer simulations to understand how these newly created ions get swept up into the flowing plasma and what role the waves play in this process. Our simulations show that within just a few rotational cycles of the ions, the interaction generates three types of waves that reach a few percent of the background magnetic field strength. These waves scatter the ions in different directions, helping the newly created ions mix efficiently with the background plasma. This process is important for understanding the complex plasma environments around active moons like Enceladus, Io, and Europa, and our results can help interpret measurements from past and future spacecraft missions to these fascinating worlds.

%
%

%


%
%
%
%

\section{Introduction}
Ion pickup at the outer planets' active moons is a fundamental process that connects local-scale surface and atmospheric activities to global-scale planetary magnetospheric dynamics. These moons create neutral gas tori around their orbits through volcanic outgassing and sublimation (e.g., Io) \cite{bagenal2020space,kivelson2004magnetospheric,roth2025mass} and water vapor plumes from subsurface oceans (e.g., Europa and Enceladus) \cite{hansen2006enceladus,waite2006cassini}. Once neutral atoms and molecules are released from these moons, they are ionized through several pathways, including photoionization by solar ultraviolet radiation, electron impact ionization from the magnetospheric electron population, and charge exchange with existing ions in the magnetosphere. These newly created ions are ``picked up'' by the planets' rotating magnetic field, interacting simultaneously with the ambient plasma via electromagnetic instabilities \cite{huddleston1997ion,huddleston2000io,russell2001dynamics,cowee2012electromagnetic}. Eventually, the newly created ions reach a kinetic equilibrium with the ambient plasma \cite{crary2000ion,cowee20071d}, transferring mass and momentum from the moons to the planetary magnetospheres. The newly picked-up ions are then transported outward through the magnetosphere via interchange instabilities and radial diffusion by strong centrifugal forces due to the rapidly rotating planets \cite{fazakerley1992drift,rymer2009cassini}. Over geological timescales, ion pickup represents a pathway for atmospheric escape that, balanced against geological supply rates, influences atmospheric evolution and potentially affects ocean stability and habitability at these worlds \cite{gronoff2020atmospheric}. Thus, the ion pickup process is a critical bridge that links the local ion production at moon atmospheres and surfaces to the global magnetospheric circulation and evolution.

\begin{figure}[tphb]
    \centering
    \includegraphics[width=0.75\linewidth]{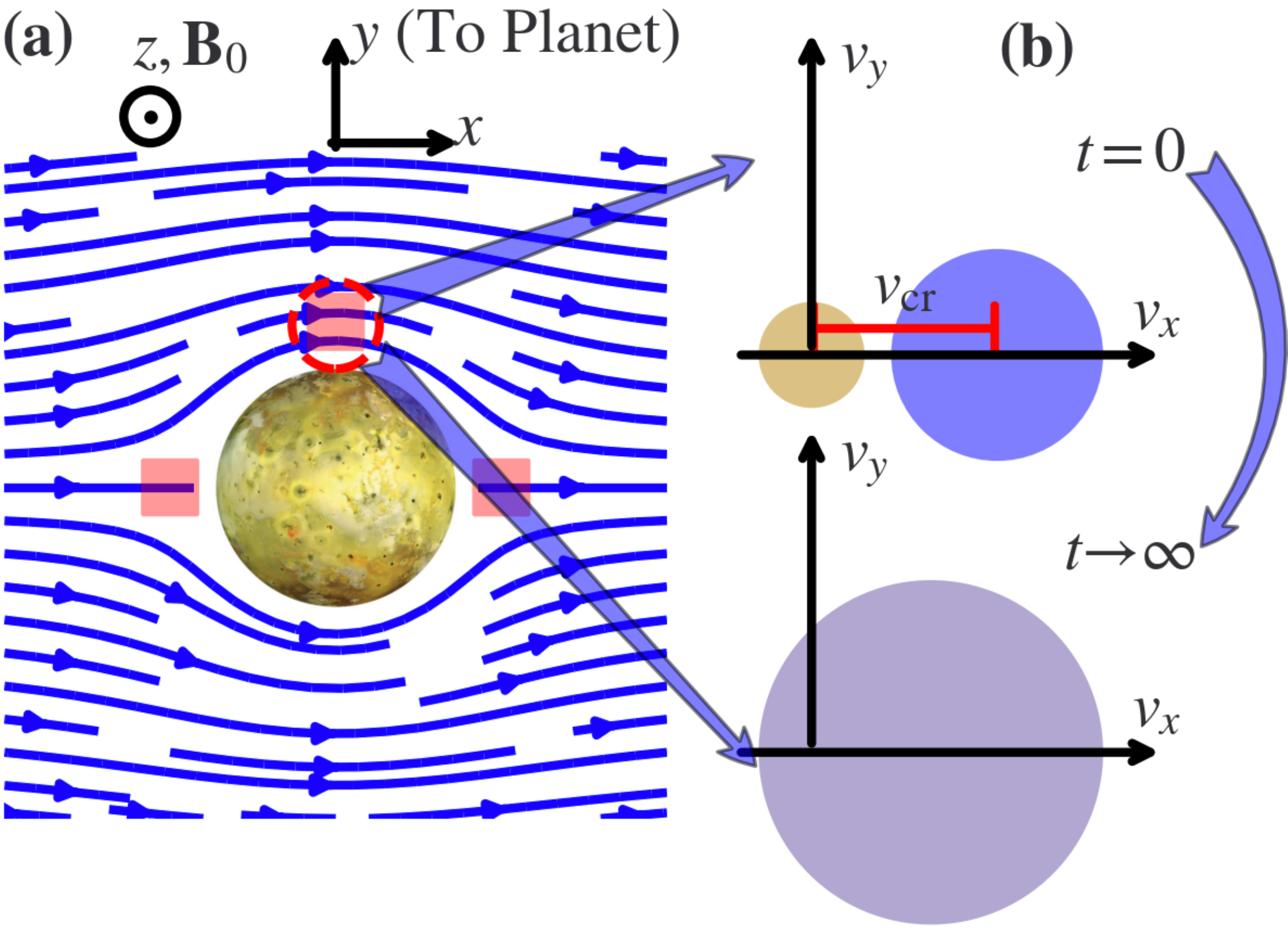}
    \caption{Sketch of the basic configuration of ion pickup at outer planets' moons. (a) Equatorial view of a moon interacting with ambient plasma corotating with the planet. Red rectangles indicate local simulation boxes. (b) Relaxation of ambient and pickup ion velocity distributions toward thermal equilibrium in the $v_x$-$v_y$ plane perpendicular to $\mathbf{B}_0$.}
    \label{fig:sketch}
\end{figure}

Figure \ref{fig:sketch} illustrates the basic configuration of ion pickup in planetary magnetospheres. In the equatorial plane, ambient plasma corotates with the planet, flowing in the $+x$ direction nearly perpendicular to the background magnetic field $\mathbf{B}_0$ (oriented along $+z$) [Figure \ref{fig:sketch}(a)]. The $+y$ direction points toward the planet, completing the orthogonal coordinate system. From the moons' reference frame, ambient ions move at a velocity determined by the difference between the corotation and orbital velocities, with corotation velocity typically dominant.\add{ Following injection, pickup ions undergo a well-documented sequence of stages: an initial nongyrotropic phase during $\mathbf{E} \times \mathbf{B}$ drift, a ring distribution after a full gyration, growth of a bispherical shell via pitch-angle scattering, and eventual relaxation to a thermalized, drift Maxwellian distribution }\cite<e.g.,>[]{coates2017pickup}\add{.}

In the center-of-mass frame, \add{consistent with this initial nongyrotropic stage, }ambient and pickup ions cluster at opposite gyrophases in velocity space, producing nongyrotropic velocity distributions that are highly unstable to electromagnetic instabilities [Figure \ref{fig:sketch}(b)]. \add{This nongyrotropic configuration reflects pickup ions sampled within a spatially localized volume close to the ionization source, where most pickup ions are ``young'' (i.e., have not yet completed a full gyration since ionization) and therefore remain bunched near a narrow range of gyrophases. }These instabilities drive the system toward equilibrium while exciting plasma waves that exhibit both transverse (perpendicular to $\mathbf{B}_0$; ion cyclotron waves) and compressional (parallel to $\mathbf{B}_0$; mirror-mode and ion Bernstein waves) magnetic field perturbations. Such wave signatures have been observed at Io and Europa by the Galileo and Juno spacecraft \cite{warnecke1997ion,russell1998magnetic,russell2000ion,blanco2001galileo,kurth2017juno,kurth2023juno,volwerk2026pick}, and at Enceladus by Cassini \cite{meeks2016comprehensive,long2022statistics,radulescu2025pick,teng2025analysis}.

More broadly, ion velocity distributions similar to those in Figure \ref{fig:sketch}(b) arise in diverse space plasma environments, including solar wind interactions with comets \cite{neugebauer1987pick,brinca1988unusual,kecskemety1989pickup,coates2004ion}, shocked solar wind in Earth's magnetosheath \cite{mckean1992mirror,cairns2001stochastic,shoji2009mirror}, and active space experiments involving ion releases \cite{papadopoulos1987collisionless,bernhardt1987observations}. Lunar-originated ions, produced by charged-particle sputtering of the lunar surface \cite{halekas2012lunar,shen2024dependence}, have also been identified as pickup ions by ARTEMIS spacecraft when the Moon resides in the subsonic magnetotail flow \cite{poppe2012artemis}, a configuration analogous to that at outer planet moons \cite{liuzzo2021investigating}. Ion pickup at outer planet moons thus provides a natural laboratory for studying plasma instabilities across multiple parameter regimes, enabling comparative analysis of these fundamentally similar physical processes. Of particular interest is the regime in which the ion source rate varies dynamically in gyrophase, as is likely the case for pickup at the outer planet moons studied here. This constitutes a distinct plasma regime for instabilities driven by nongyrotropic distributions, potentially differing significantly from the more conventional cases of gyrotropic temperature-anisotropic or ring-beam ion distributions.

Using kinetic simulations, we examine three key aspects of nongyrotropic ion distribution relaxation in the pickup process at outer planet moons: the evolution of ion velocity distributions (Section~\ref{sec:ion-dists}); the properties of the resulting unstable wave modes (Sections~\ref{sec:wave-excitation} and~\ref{sec:density}); and the resonances and velocity-space structures that drive wave growth and the simultaneous relaxation of nongyrotropic distributions (Section~\ref{sec:fpcorr}). We summarize our results in Section~\ref{sec:summary}.

\section{Computational setup}
We employ the hybrid-VPIC code \cite{bowers2008ultrahigh,le2023hybrid} to investigate the ion pickup process, modeling ions as particles and electrons as a massless fluid. The simulation uses a two-dimensional configuration space $(x, z)$ and three-dimensional velocity space $(v_x, v_y, v_z)$, with the background magnetic field directed along $z$ and periodic boundary conditions applied in both directions. To isolate the basic pickup mechanism, we consider O$^+$ ions for both the ambient and pickup populations, representative of \change{the dominant}{a dominant} ion species in Io-Jupiter interactions\add{ }\cite{kim2020survey}.\add{ Protons are not included: restricting the simulation to a single, equal-mass species isolates the instability driven by the relative drift and gyrophase bunching between the ambient and pickup populations, without additional wave modes (e.g., proton cyclotron waves) that a lighter species would introduce. Since a single ion species is used, we adopt the oxygen ion mass $m_O$ as the mass unit ($m_O = 1$); the ion inertial length $d$, Alfv\'en velocity $v_{\mathrm{A}}$, and gyrofrequency $\omega_c$ used throughout are defined accordingly in terms of $m_O$ and the reference density $n_0$.} 

The simulation parameters follow \citeA{huddleston1999mirror}: a background magnetic field $B_0 = 1720$~nT and total plasma density $n_0 = 5000$~cm$^{-3}$, which yield an Alfv\'en velocity $v_{\mathrm{A}} = 133$~km/s for O$^+$. The domain size is $L_x = L_z = 128 d$, where $d$ is the oxygen ion inertial length, discretized with $N_x = N_z = 512$ grid points and $200$ particles per cell. The time step $\Delta t = (1/1600) \omega_c^{-1}$ satisfies the Courant condition $\Delta t \omega_c < (\Delta x / d)^2 / (\sqrt{2} \pi)$, where $\omega_c$ is the oxygen ion gyrofrequency and $\Delta x$ is the cell size. \add{This quadratic (rather than linear) dependence on $\Delta x$ arises because the fastest propagating mode in hybrid simulations is the whistler wave rather than the light wave. Requiring the wave to propagate no more than one cell per time step then yields the parabolic stability condition $\Delta t \propto (\Delta x)^2$ }\cite{pritchett2000particle-in-cell-8dc}\add{.}

The electron temperature is set to $T_e = 5$\,eV, corresponding to a thermal velocity $940$\,km/s, which greatly exceeds all other characteristic velocities (ion thermal velocities, Alfv\'en velocity, and corotation velocity). This ordering justifies the use of an isothermal equation of state ($T_e = \mathrm{constant}$) and a fluid description for electrons, as electron kinetics are less critical for ion pickup dynamics.

The ambient corotating and pickup ion populations have equal densities of $n_{\mathrm{cr}}= n_{\mathrm{pu}} = 0.5 n_0$ \cite{bagenal1994empirical,frank1996plasma,huddleston1999mirror}. The corotating ions are initialized as a Maxwellian with temperature $T_{\mathrm{cr}}=20$\,eV (plasma beta $\beta_{\mathrm{cr}}=6.8 \times 10^{-3}$), while the pickup ions, initially stationary in the moon's frame, have temperature $T_{\mathrm{pu}}=1$\,eV ($\beta_{\mathrm{pu}}=3.4 \times 10^{-4}$). The two populations are separated by the corotation velocity $v_{\mathrm{cr}} = 0.9 v_{\mathrm{A}} = 120$\,km/s in the $x$ direction, providing the free energy that drives electromagnetic instabilities. The moon's orbital velocity is negligible compared to the corotation velocity and is therefore set to zero.\add{ Our simulation is thus conducted in the moon's rest frame. Because the center of mass moves at $v_{\mathrm{cm}} = 0.5\,v_{\mathrm{cr}} = 0.45\,v_{\mathrm{A}}$ along $\hat{x}$ in this frame, a convection electric field $E_{y0} = v_{\mathrm{cm}} B_0/c = 0.45\,(v_{\mathrm{A}} B_0/c)$ is present self-consistently throughout the simulation.}

Unlike conventional pickup studies that assume ring velocity distributions formed by rapid ion gyrotropization, we explicitly consider potentially nongyrotropic distributions ($\partial_\phi f \neq 0$, where $\phi$ is the gyrophase and $f$ is the distribution function) that can arise from variable or impulsive ionization rates. This approach captures the regime in which ion gyration timescales are not necessarily much shorter than wave growth or thermalization timescales, thereby revealing additional pickup dynamics beyond the idealized ring-distribution picture.

\section{Evolution of ion velocity distributions}\label{sec:ion-dists}
\begin{figure}[h]
    \centering
    \includegraphics[width=\linewidth]{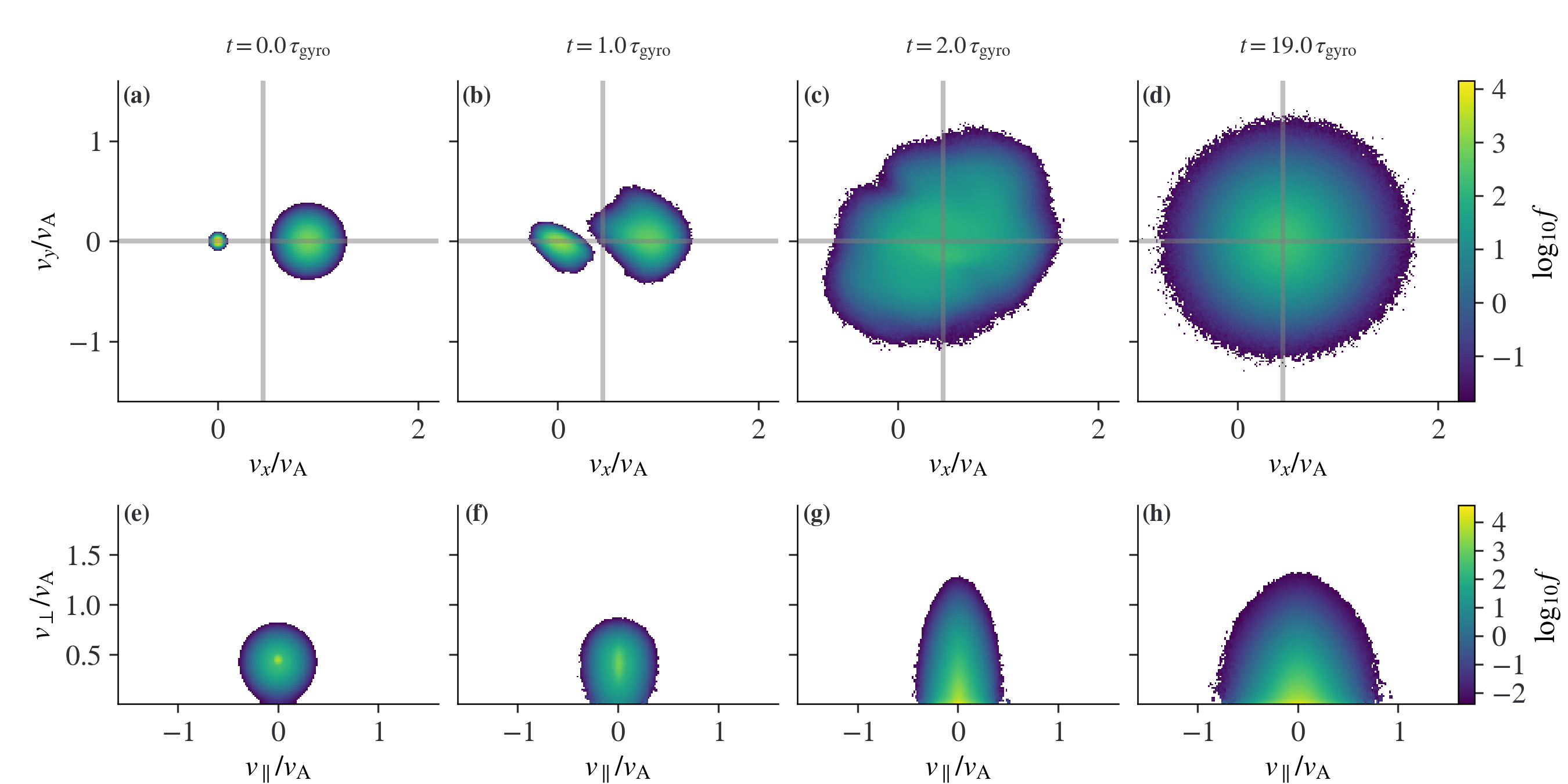}
    \caption{Reduced ion velocity distributions at four representative times. Each column corresponds to $t/\tau_{\mathrm{gyro}} = 0.0$, $1.0$, $2.0$, and $19.0$. (a--d) Reduced distributions $\int \mathrm{d}v_z\, F(v_x, v_y, v_z)$ in the plane $(v_x, v_y)$ perpendicular to the ambient magnetic field $B_0$ ($F$ being the full ion velocity distribution function).\add{ The velocity distributions in these four panels are in the moon's rest frame.} The gray cross marks the center-of-mass velocity, located at $(v_x = 0.45 v_{\mathrm{A}}, v_y = 0)$. (e--h) Reduced distributions $\int \mathrm{d}\phi\, F(v_\parallel, v_\perp, \phi)$ in the plane $(v_\parallel, v_\perp)$.\add{ The velocity distributions in these four panels are in the center-of-mass frame.} The center-of-mass velocity appears at $(v_\parallel = 0, v_\perp = 0)$.}
    \label{fig:fdists}
\end{figure}

We show four representative snapshots of reduced ion velocity distributions in Figure \ref{fig:fdists}. Initially, pickup ions are centered at the origin $(v_x = 0, v_y = 0)$ with a small thermal spread in the \change{Moon}{moon} frame, while ambient corotating ions are centered at the corotation velocity $(v_x = 0.9 v_\mathrm{A}, v_y = 0)$ [Figure \ref{fig:fdists}(a)]. Since the two populations have equal densities ($n_\mathrm{pu} = n_\mathrm{cr} = 0.5\,n_0$), the total momentum is conserved at the center-of-mass velocity $v_\mathrm{cm} = (0 + 0.9\,v_\mathrm{A})/2 = 0.45\,v_\mathrm{A}$, located at $(v_x = 0.45\,v_\mathrm{A}, v_y = 0)$. When viewed in $(v_\parallel, v_\perp)$ space in the center-of-mass frame [Figure \ref{fig:fdists}(e)], the gyrophase is integrated out, revealing pickup ions as a cold ring centered at $(v_\parallel = 0, v_\perp = 0.45 v_{\mathrm{A}})$, embedded within a more diffuse ring of ambient ions centered at the same location. In the absence of electromagnetic instabilities, momentum conservation requires that both populations gyrate indefinitely around $v_\mathrm{cm}$ in the $(v_x, v_y)$ plane without gyrophase mixing.

After one gyroperiod ($t = 1.0 \tau_{\mathrm{gyro}}$, where $\tau_{\mathrm{gyro}}$ is the oxygen ion gyroperiod), pickup and ambient ions gyrate approximately $360^\circ$ about the center-of-mass velocity, with moderate modifications to the Maxwellian shape in velocity space [Figures \ref{fig:fdists}(b) and \ref{fig:fdists}(f)]. After two gyroperiods ($t = 2.0 \tau_{\mathrm{gyro}}$), both populations are dragged toward the center-of-mass velocity [Figures \ref{fig:fdists}(c) and \ref{fig:fdists}(g)], with significant thermalization in $v_\perp$. Notably, the time scale for thermalization in gyrophase $\phi$ is comparable to that for drift and thermalization in $v_{\perp}$, contrary to the conventional assumption that gyrophase mixing occurs on much shorter timescales than scattering in $v_{\perp}$. Approaching the end of the simulation ($t = 19\,\tau_{\mathrm{gyro}}$), pickup and ambient ions merge into a single fully thermalized Maxwellian [Figure \ref{fig:fdists}(d)], drifting at exactly $v_\mathrm{cm} = 0.45\,v_\mathrm{A}$ in the $+x$ direction, confirming that total momentum is conserved throughout the interaction. The free energy released by the instability drives thermalization in both the perpendicular and parallel directions [Figure \ref{fig:fdists}(h)], but cannot alter the bulk drift of the combined system.

\begin{figure}[h]
    \centering
    \includegraphics[width=\linewidth]{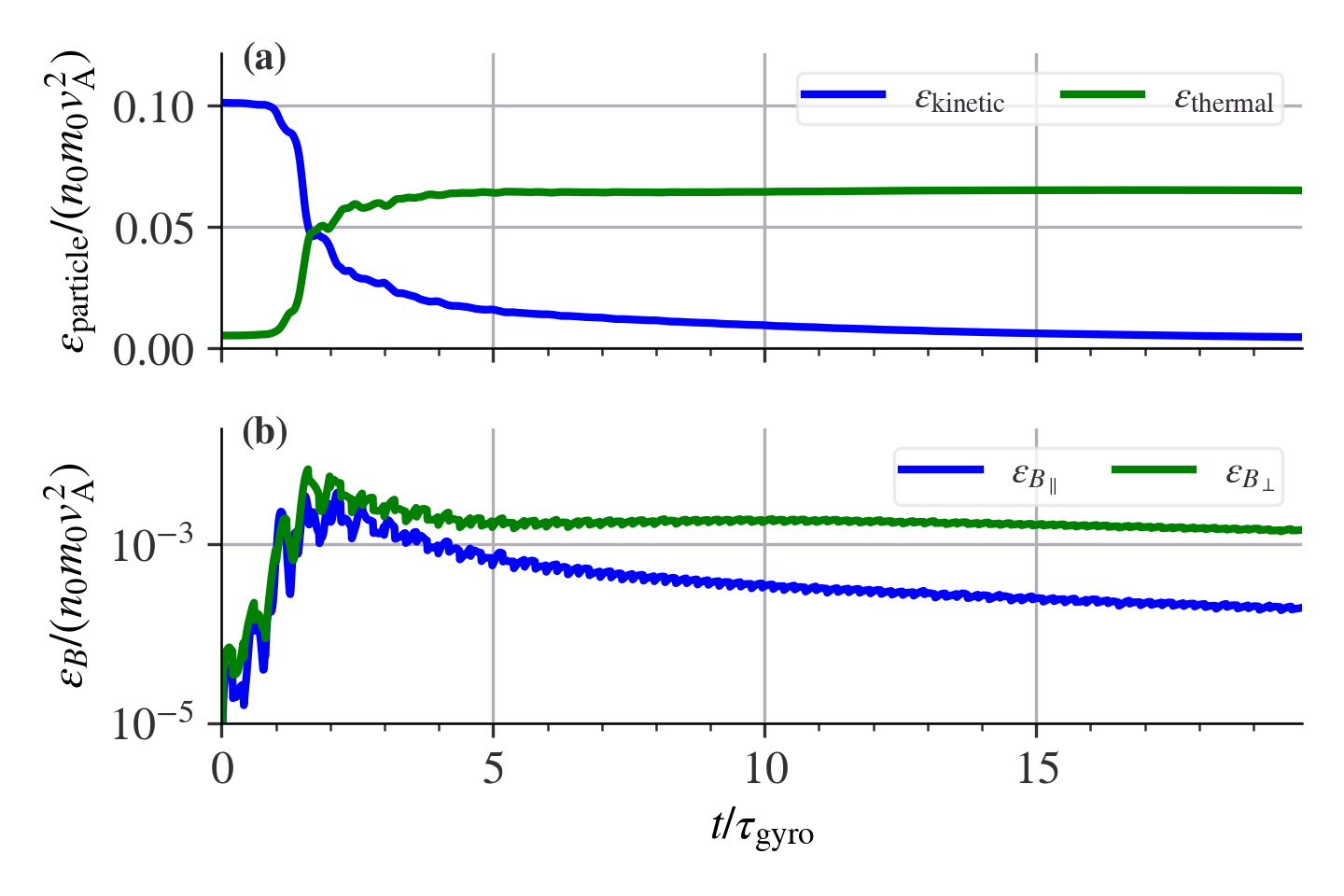}
    \caption{Temporal evolution of particle and magnetic field energy densities. (a) Total bulk kinetic energy density (blue) and thermal energy density (green) of both ion species. The bulk kinetic energy is evaluated in the center-of-mass frame, so its asymptotic value approaches zero as $t \to \infty$. The final thermal energy density does not reach the initial bulk kinetic energy density \remove{due to finite dissipation from numerical resistivity (which is necessary to stabilize the hybrid-kinetic algorithm)}\add{because, although physical resistivity $\eta$ and hyper-resistivity $\eta_H$ are both set to zero, numerical dissipation arises inevitably from the linear extrapolation of the ion current density in Ohm's law and smoothing of the magnetic field} \cite{le2023hybrid}\add{, which is necessary to stabilize the explicit hybrid-kinetic algorithm against grid-scale whistler oscillations in the nonlinear regime}. (b) Compressional ($B_\parallel$, blue) and transverse ($B_\perp$, green) magnetic field energy densities. The dependence of wave growth on gyrophase is clearly visible in the first two gyroperiods.}
    \label{fig:energy-budget}
\end{figure}

The thermalization of the nongyrotropic ion velocity distribution must be mediated by self-generated electromagnetic waves. These waves draw their free energy from the relative drift between pickup ions and ambient corotating ions perpendicular to the background magnetic field. Figure~\ref{fig:energy-budget} shows the temporal evolution of particle and field energies. For convenience, we compute the total bulk kinetic energy density of both ion \change{species}{populations} in the center-of-mass frame. In this frame, the final bulk kinetic energy density vanishes, because all ions ultimately converge to the center-of-mass velocity. The initial bulk kinetic energy density is
\begin{linenomath*}
    \begin{align}
        \varepsilon_{\mathrm{kinetic}} &= \frac{1}{2} n_{\mathrm{cr}} m_0 (v_{\mathrm{cr}} - v_{\mathrm{cm}})^2 + \frac{1}{2} n_{\mathrm{pu}} m_0 (0 - v_{\mathrm{cm}})^2 \nonumber \\
        &= \frac{1}{2} n_0 m_0 v_{\mathrm{A}}^2 M_{\mathrm{A}}^2 \eta_{\mathrm{cr}} (1 - \eta_{\mathrm{cr}}) ,
    \end{align}
\end{linenomath*}
where $m_0$ is the oxygen mass, $v_{\mathrm{cm}}=\eta_{\mathrm{cr}} v_{\mathrm{cr}} + \eta_{\mathrm{pu}} \cdot 0$ is the center-of-mass velocity, $\eta_{\mathrm{cr}} = n_{\mathrm{cr}}/n_0$ is the ambient-ion fraction, $\eta_{\mathrm{pu}} = n_{\mathrm{pu}}/n_0$ is the pickup-ion fraction, and $M_{\mathrm{A}} = v_{\mathrm{cr}}/v_{\mathrm{A}}$ is the Alfv\'en Mach number of the corotating ambient flow. Within two ion gyroperiods ($0 < t/\tau_{\mathrm{gyro}} < 2$), both transverse ($B_\perp$) and compressional ($B_\parallel$) magnetic perturbations grow exponentially and then saturate. These waves correspond to the electromagnetic ion-cyclotron (EMIC; also know as ion cyclotron waves or ICWs), mirror-mode, and ion Bernstein waves (Section~\ref{sec:wave-excitation}). Their large amplitudes ($B_\perp/B_0 \approx 0.03$ and $B_\parallel/B_0 \approx 0.03$) efficiently scatter both ion populations in velocity space $(v_\parallel, v_\perp, \phi)$ (Section~\ref{sec:fpcorr}), leading to rapid thermalization and the eventual merging of the two ion populations. Although the wave energy remains small compared to the particle energy, the waves provide the mechanism for converting bulk kinetic energy into thermal energy. As shown in Figure~\ref{fig:energy-budget}, the energy transfer rates, $\vert \mathrm{d}\varepsilon_{\mathrm{kinetic}}/\mathrm{d}t \vert$ and $\vert \mathrm{d}\varepsilon_{\mathrm{thermal}}/\mathrm{d}t \vert$, peak when the wave amplitudes are largest at \change{$t = 2.0 \tau_{\mathrm{gyro}}$}{$1 < t / \tau_{\mathrm{gyro}} < 2$}.

In the limit where the thermal speed of either ion population exceeds the initial drift separation (i.e., the corotation velocity), electromagnetic waves are only weakly excited because the overall ion distribution is close to kinetic equilibrium. To quantify this, we evaluate the initial total thermal energy density,
\begin{linenomath*}
    \begin{align}
        \varepsilon_{\mathrm{thermal}} &= \frac{3}{2} n_{\mathrm{cr}} m_0 v_{\mathrm{Tcr}}^2 + \frac{3}{2} n_{\mathrm{pu}} m_0 v_{\mathrm{Tpu}}^2 \nonumber \\
        &= \frac{3}{4} n_0 m_0 v_{\mathrm{A}}^2 \left[\eta_{\mathrm{cr}} \beta_{\mathrm{cr}} + (1-\eta_{\mathrm{cr}}) \beta_{\mathrm{pu}}\right] \nonumber \\
        &\approx \frac{3}{4} n_0 m_0 v_{\mathrm{A}}^2 \eta_{\mathrm{cr}} \beta_{\mathrm{cr}} ,
    \end{align}
\end{linenomath*}
where the final approximation assumes cold pickup ions, $\eta_{\mathrm{cr}} \beta_{\mathrm{cr}} \gg (1-\eta_{\mathrm{cr}}) \beta_{\mathrm{pu}}$. A necessary but not sufficient condition for wave excitation is $\varepsilon_{\mathrm{kinetic}}/\varepsilon_{\mathrm{thermal}} > 1$, which gives
\begin{linenomath*}
    \begin{align}
        \frac{2 M_{\mathrm{A}}^2 (1 - \eta_{\mathrm{cr}})}{3 \beta_{\mathrm{cr}}} > 1 . \label{eq:threshold-pickup}
    \end{align}
\end{linenomath*}
This resembles the familiar instability thresholds for EMIC and mirror-mode waves in the solar wind and Earth's magnetosheath \cite{Bale09,shoji2009mirror},
\begin{linenomath*}
    \begin{align}
        \frac{T_\perp}{T_\parallel} > 1 + \frac{1}{\beta_{\perp}} , \label{eq:threshold-bimax}
    \end{align}
\end{linenomath*}
where $T_\perp$ and $T_\parallel$ are the perpendicular and parallel temperatures of a bi-Maxwellian ion distribution, and $\beta_\perp = n_0 T_\perp/(B_0^2/8\pi)$. Equation~\eqref{eq:threshold-pickup} parallels Equation~\eqref{eq:threshold-bimax} in that the corotating velocity acts as an effective perpendicular thermal speed. The precise threshold on the right-hand side of Equation~\eqref{eq:threshold-pickup}, analogous to the $1/\beta_\perp$ dependence in Equation~\eqref{eq:threshold-bimax}, can be determined through linear kinetic instability analysis or parameter scans in hybrid-kinetic simulations, which we will address in a future study.

\section{Identification of EMIC, mirror-mode, and ion Bernstein waves}\label{sec:wave-excitation}
We perform a Fourier analysis to identify the plasma waves excited by the perpendicular streaming between ambient and pickup ions. The magnetic field is first decomposed in space as
\begin{linenomath*}
\begin{align}
    \mathbf{B}(\mathbf{x}, t) = \int \frac{\mathrm{d}^3 k}{(2\pi)^3} \exp(i \mathbf{k}\cdot\mathbf{x})\, \mathbf{B}_{\mathbf{k}}(t) .
\end{align}
\end{linenomath*}
The corresponding dispersion relation in $(\mathbf{k}, \omega)$ space is then obtained by Fourier-transforming $\mathbf{B}_{\mathbf{k}}(t)$ in time:
\begin{linenomath*}
\begin{align}
    \mathbf{B}_{\mathbf{k}}(t) = \int \mathrm{d}\omega\, \exp(-i\omega t)\, \mathbf{B}_{\mathbf{k}, \omega} .
\end{align}
\end{linenomath*}

We express the fluctuating magnetic field components in terms of circular polarization modes about $B_0$. The perpendicular components are decomposed into right-handed $(r)$ and left-handed $(l)$ circular polarizations, while the parallel component $(\parallel)$ remains unchanged. This representation is particularly useful because the linear eigenmodes in the parallel propagation limit consist of pure right-handed, left-handed, or parallel polarizations:
\begin{linenomath*}
    \begin{align}\label{eq:polarization-def}
        B_{\mathbf{k}}^l = \left( \frac{B_x + i B_y}{\sqrt{2}} \right)_{\mathbf{k}}, \quad B_{\mathbf{k}}^r = \left( \frac{B_x - i B_y}{\sqrt{2}} \right)_{\mathbf{k}}, \quad
        B_{\mathbf{k}}^\parallel = \left( B_z \right)_{\mathbf{k}},
    \end{align}
\end{linenomath*}
with corresponding unit vectors
\begin{linenomath*}
    \begin{align}
        \mathbf{e}_l = \frac{1}{\sqrt{2}}\left( \mathbf{e}_x - i \mathbf{e}_y \right), \quad \mathbf{e}_r = \frac{1}{\sqrt{2}}\left( \mathbf{e}_x + i \mathbf{e}_y \right), \quad \mathbf{e}_{\parallel} = \mathbf{e}_z.
    \end{align}
\end{linenomath*}

\begin{figure}[h]
    \centering
    \includegraphics[width=\linewidth]{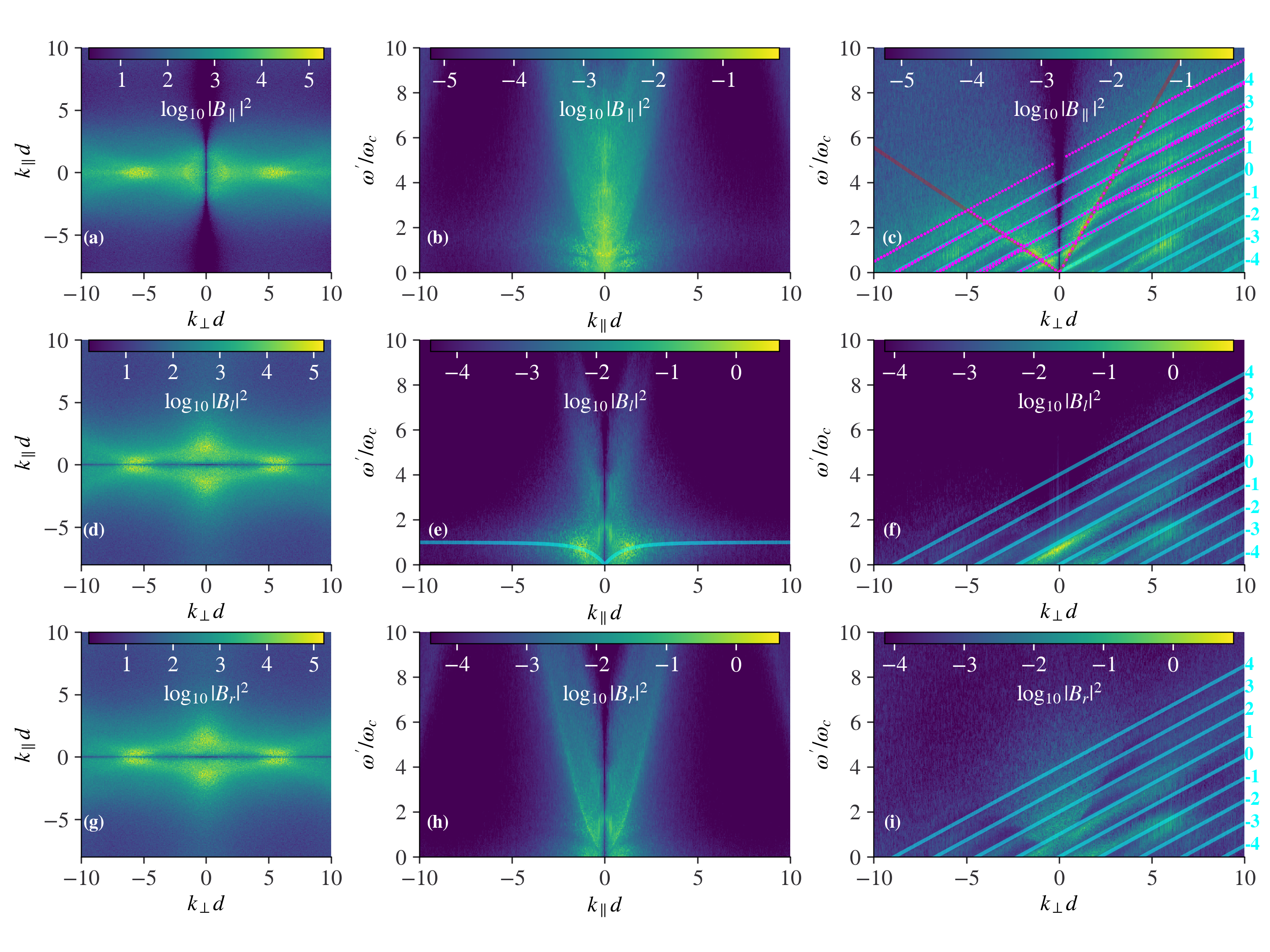}
    \caption{Fourier analysis of magnetic field perturbations. The three rows (top to bottom) display results for compressional, left-hand polarized, and right-hand polarized magnetic field components, respectively. The three columns (left to right) display magnetic power in $(k_\perp, k_\parallel)$, $(k_\parallel, \omega')$, and $(k_\perp, \omega')$ space, respectively.\add{ The magenta dots in panel (c) represent the dispersion relation of the extraordinary mode (including ion Bernstein waves) propagating perpendicular to the background magnetic field. The two phase velocities $v_{\mathrm{cm}} \pm v_f$ are marked as red lines. The dispersion relation for the mirror mode is simply $\omega = \omega' - k_\perp v_{\mathrm{cm}} = 0$, since this mode is stationary in the plasma rest frame.} The \change{white}{cyan} curve in panel (e) represents the cold plasma dispersion relation for EMIC waves. \change{White}{Cyan} lines in panels (c), (f), and (i) represent ion gyrofrequency harmonics in the center-of-mass frame, $\omega / \omega_c = (\omega' - k_\perp v_{\mathrm{cm}}) / \omega_c = 0, \pm 1, \pm 2, \pm 3, \pm 4$. Note that $\omega$ and $\omega'$ denote the wave frequencies in the center-of-mass and \change{Moon's}{moon's} frames, respectively.}  
    \label{fig:disps}
\end{figure}

\begin{figure}[h]
    \centering
    \includegraphics[width=\linewidth]{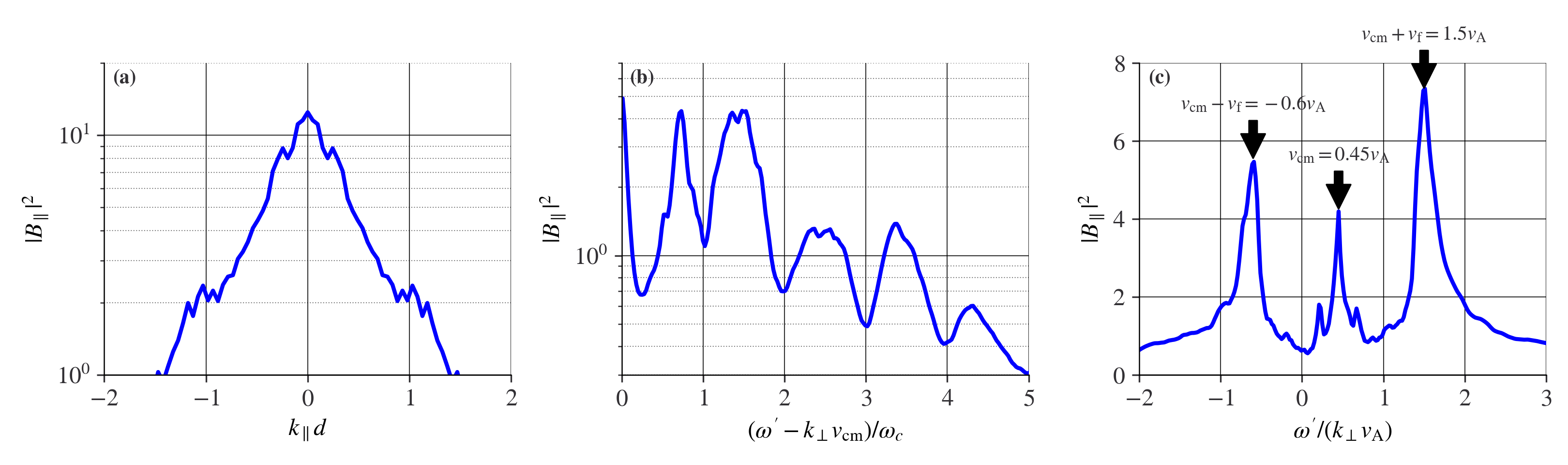}
    \caption{Characteristics of mirror-mode and ion Bernstein waves. (a) Spread of compressional wave power around $k_\parallel = 0$. (b) Frequency spectrum of $\vert B_\parallel \vert^2$ in the center-of-mass frame showing the mirror mode at $\omega = \omega' - k_\perp v_{\mathrm{cm}} = 0$, ion Bernstein modes between ion gyrofrequency harmonics, and the EMIC mode at $\omega / \omega_c = (\omega' - k_\perp v_{\mathrm{cm}}) / \omega_c = 0.7$. (c) $\vert B_\parallel \vert^2$ as a function of perpendicular phase velocity, showing the mirror mode at $\omega' / k_\perp = v_{\mathrm{cm}} = 0.45 v_{\mathrm{A}}$ and ion Bernstein waves at \change{$\omega' / k_\perp = v_{\mathrm{cm}} \pm v_{\mathrm{f}} = 0.6 v_{\mathrm{A}},\, 1.5 v_{\mathrm{A}}$}{$\omega' / k_\perp = v_{\mathrm{cm}} \pm v_{\mathrm{f}} = -0.6 v_{\mathrm{A}},\, 1.5 v_{\mathrm{A}}$}.}
    \label{fig:disps-bpara}
\end{figure}

\begin{figure}[h]
    \centering
    \includegraphics[width=\linewidth]{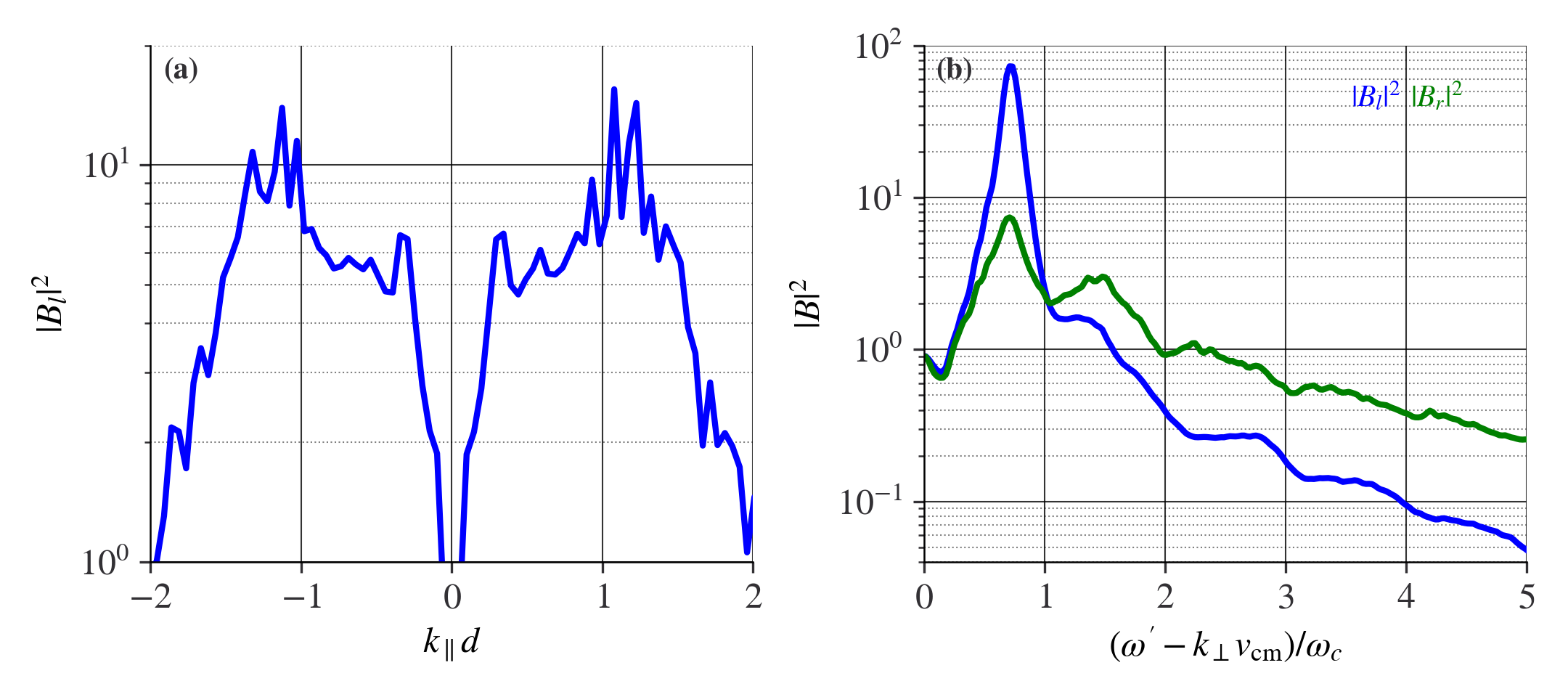}
    \caption{Characteristics of EMIC waves. (a) Peak of $\vert B_l \vert^2$ at $k_\parallel d = \pm 1.1$. Note that the power in $\vert B_l \vert^2$ near $k_\parallel = 0$ is predominantly contributed by mirror-mode and ion Bernstein waves. (b) Frequency spectra of $\vert B_l \vert^2$ (blue) and $\vert B_r \vert^2$ (green) in the center-of-mass frame showing EMIC waves at $\omega / \omega_c = (\omega' - k_\perp v_{\mathrm{cm}}) / \omega_c = 0.7$.}
    \label{fig:disps-bleft}
\end{figure}

\add{Figures~}\ref{fig:disps}\add{, }\ref{fig:disps-bpara}\add{, and }\ref{fig:disps-bleft}\add{ show the Fourier analysis of magnetic field perturbations. The three rows (top to bottom) of Figure~}\ref{fig:disps}\add{ display results for the compressional ($B_\parallel$), left-hand polarized ($B_l$), and right-hand polarized ($B_r$) magnetic field components, respectively. The three columns (left to right) display magnetic power in $(k_\perp, k_\parallel)$, $(k_\parallel, \omega')$, and $(k_\perp, \omega')$ space, respectively, where $\omega'$ represents the wave frequency in the moon's (simulation) frame and is related to the wave frequency $\omega$ in the center-of-mass frame via $\omega' = \omega + k_\perp v_{\mathrm{cm}}$. Figures~}\ref{fig:disps-bpara}\add{ and }\ref{fig:disps-bleft}\add{ then present one-dimensional cuts of this power spectrum, isolating the parallel wavenumber, frequency, and phase-velocity dependence, respectively.}

\remove{Figures }\remove{(a-c) show the Fourier analysis of compressional magnetic field perturbations ($B_\parallel$). }The power in $B_\parallel$ is concentrated in nearly perpendicular propagation to $B_0$ [$k_\perp \gg k_\parallel \sim 0$, Figures~\ref{fig:disps}(a) and \ref{fig:disps-bpara}(a)], with peaks at \change{$k_\perp d = 1.5$}{$k_\perp d = \pm 1.5$} and \change{$k_\perp d = 5.5$}{$k_\perp d = \pm 5.5$}. In $(k_\perp, \omega')$ space, the power spectrum $\vert B_\parallel \vert^2$ is organized by the dispersion relations of mirror-mode and ion Bernstein waves [Figure \ref{fig:disps}(c)\add{ and }\ref{app:ibw}]. Here $\omega'$ represents the wave frequency in the \change{Moon's}{moon's} (simulation) frame and is related to the wave frequency $\omega$ in the center-of-mass frame via $\omega' = \omega + k_\perp v_{\mathrm{cm}}$. Mirror-mode waves peak at zero frequency in the center-of-mass frame [Figure~\ref{fig:disps-bpara}(b)], corresponding to propagation at $\omega' / k_\perp = v_{\mathrm{cm}} = 0.45 v_{\mathrm{A}}$ in the moon's frame [Figure~\ref{fig:disps-bpara}(c)]. Ion Bernstein waves, on the other hand, peak between harmonics of the ion gyrofrequency in the center-of-mass frame [Figures~\ref{fig:disps}(c) and \ref{fig:disps-bpara}(b)]. Most of the ion Bernstein wave power is concentrated at the fast magnetosonic velocity $v_{\mathrm{f}}$ in the center-of-mass frame [Figure~\ref{fig:disps-bpara}(c)], corresponding to $\omega / k_\perp = \omega' / k_\perp - v_{\mathrm{cm}} = \pm v_{\mathrm{f}} = \pm 1.05 v_{\mathrm{A}}$. This is the origin of the power bands along these two phase velocities in Figure \ref{fig:disps}(c). Finally, a distinct power band appears at $\omega/\omega_c = (\omega' - k_\perp v_{\mathrm{cm}})/\omega_c \lesssim 1$ in both $|B_\parallel|^2$ and $|B_l|^2$ [Figures~\ref{fig:disps}(c) and \ref{fig:disps}(f)], manifesting as a linear band of power just below the $\omega' - k_\perp v_{\mathrm{cm}} = \omega_c$ line and a spectral peak at $\omega'/\omega_c < 1$ [Figure~\ref{fig:disps}(e)]. This feature corresponds to EMIC waves, which in the center-of-mass frame peak just below the ion cyclotron frequency; we discuss these further below. 

\remove{Figures (d-f) and (g-i) show the Fourier analysis of left-hand ($B_l$) and right-hand ($B_r$) polarized magnetic field perturbations, respectively. }In wavenumber space $(k_\perp, k_\parallel)$, $B_l$ and $B_r$ are indistinguishable owing to their formalism [Equation \eqref{eq:polarization-def}], but they are fully distinguished in $(k_\parallel, \omega')$ or $(k_\perp, \omega')$ space, since the different polarizations exhibit growth at different frequencies. EMIC waves appear predominantly in $\vert B_l \vert^2$ at $k_\parallel d = \pm 1.1$ and $\omega / \omega_c = 0.7$, with a small fraction of their power appearing in $\vert B_\parallel \vert^2$ and $\vert B_r \vert^2$ due to their finite $k_\perp$ [Figure~\ref{fig:disps-bpara}(b), as discussed previously for $|B_\parallel|^2$, and Figure~\ref{fig:disps-bleft}(b) for the two transverse polarizations]. Similarly, the peaks in $\vert B_l \vert^2$ and $\vert B_r \vert^2$ at $k_\perp d = 5.5$ arise from the small but finite $k_\parallel$ of ion Bernstein waves [Figures~\ref{fig:disps}(d), \ref{fig:disps}(g) and \ref{fig:disps-bpara}(a)].

Comparing the power spectral densities of compressional and transverse magnetic fields [Figures \ref{fig:disps-bpara}(b) and \ref{fig:disps-bleft}(b)], we find that transverse perturbations exhibit a relatively higher peak concentrated around $\omega = 0.7 \omega_c$ (predominantly from EMIC waves), whereas compressional perturbations are lower in magnitude but spread over a broader frequency range. Their total energies at saturation, however, are comparable [Figure \ref{fig:energy-budget}(b)], although compressional perturbations decay faster and to lower amplitudes after saturation.

\begin{figure}[h]
    \centering
    \includegraphics[width=\linewidth]{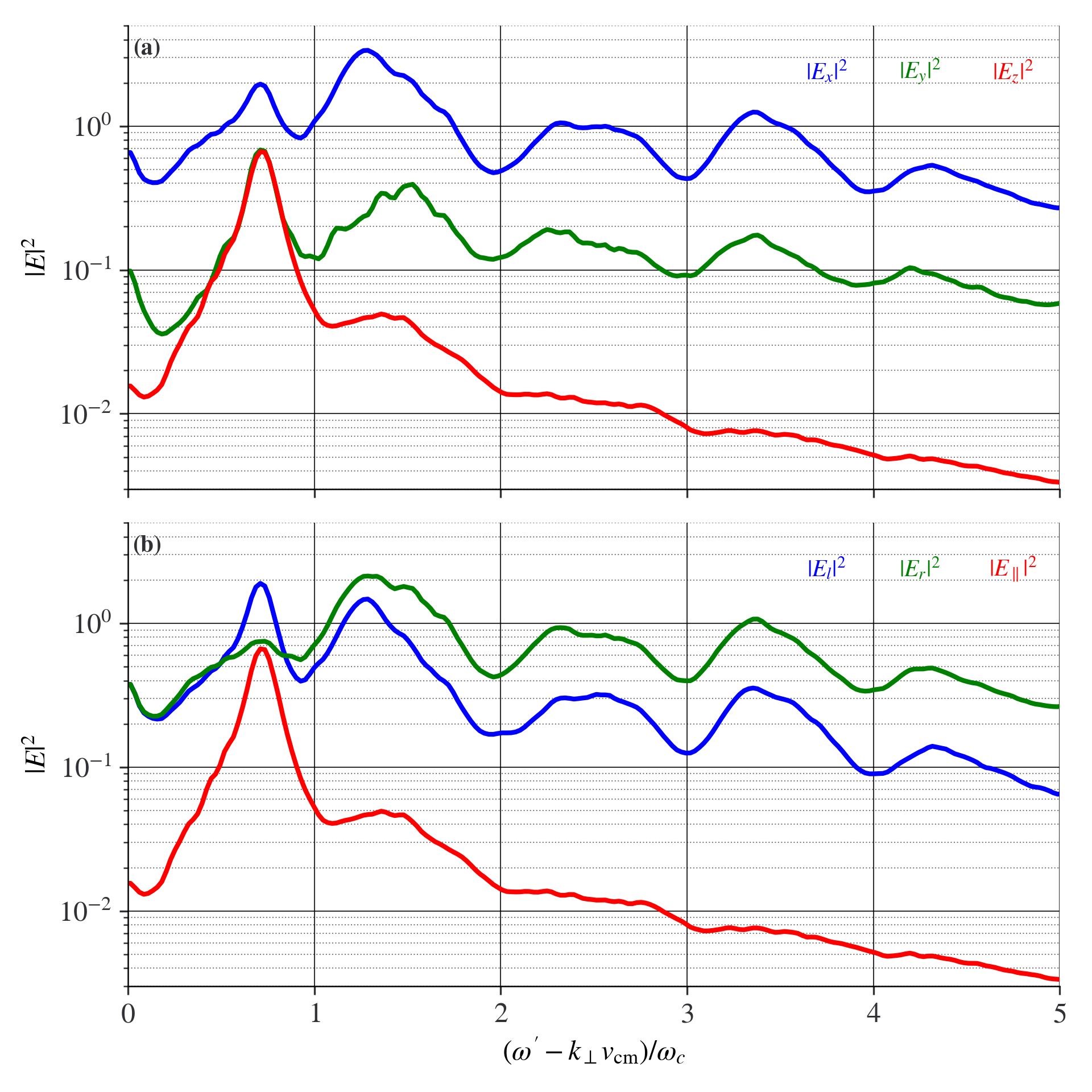}
    \caption{\add{Frequency spectra of electric fields in the center-of-mass frame. (a) Frequency spectra in Cartesian coordinates: $\vert E_x \vert^2$ (blue), $\vert E_y \vert^2$ (green) and $\vert E_z \vert^2$ (red). (b) }Frequency spectra \change{of}{in polarization-based coordinates:} $\vert E_l \vert^2$ (blue), $\vert E_r \vert^2$ (green) and $\vert E_\parallel \vert^2$ (red)\remove{ in the center-of-mass frame}.}
    \label{fig:disps-elines}
\end{figure}

Understanding the polarization of electric field perturbations is important since they directly control energy transfer between fields and particles. For mirror-mode and ion Bernstein waves, \change{the dominant electric field component is $E_y = (\omega / k_\perp c) B_\parallel$, which decomposes into $E_l$ and $E_r$ with equal magnitude}{$E_x$ dominates over $E_y$ [Figure }\ref{fig:disps-elines}\add{(a)], indicating that the electrostatic component dominates over the electromagnetic component}. Due to finite $k_\parallel$ [Figure \ref{fig:disps-bpara}(a)], ion Bernstein waves become elliptically polarized with the magnitude of $E_r$ larger than $E_l$ [Figure \ref{fig:disps-elines}\add{(b)}], consistent with the relative magnitudes of $B_r$ and $B_l$ at the corresponding frequencies [Figure \ref{fig:disps-bleft}(b)]. For quasi-parallel propagating EMIC waves around $\omega / \omega_c = 0.7$, $E_l$ is the dominant component. EMIC waves exhibit significant parallel electric fields, likely arising from a combination of finite $k_\perp$ and mode coupling driven by nongyrotropic distributions, as discussed in Section \ref{sec:fpcorr}.


\section{Density perturbations associated with different wave modes}\label{sec:density}
\begin{figure}[tphb]
    \centering
    \includegraphics[width=\linewidth]{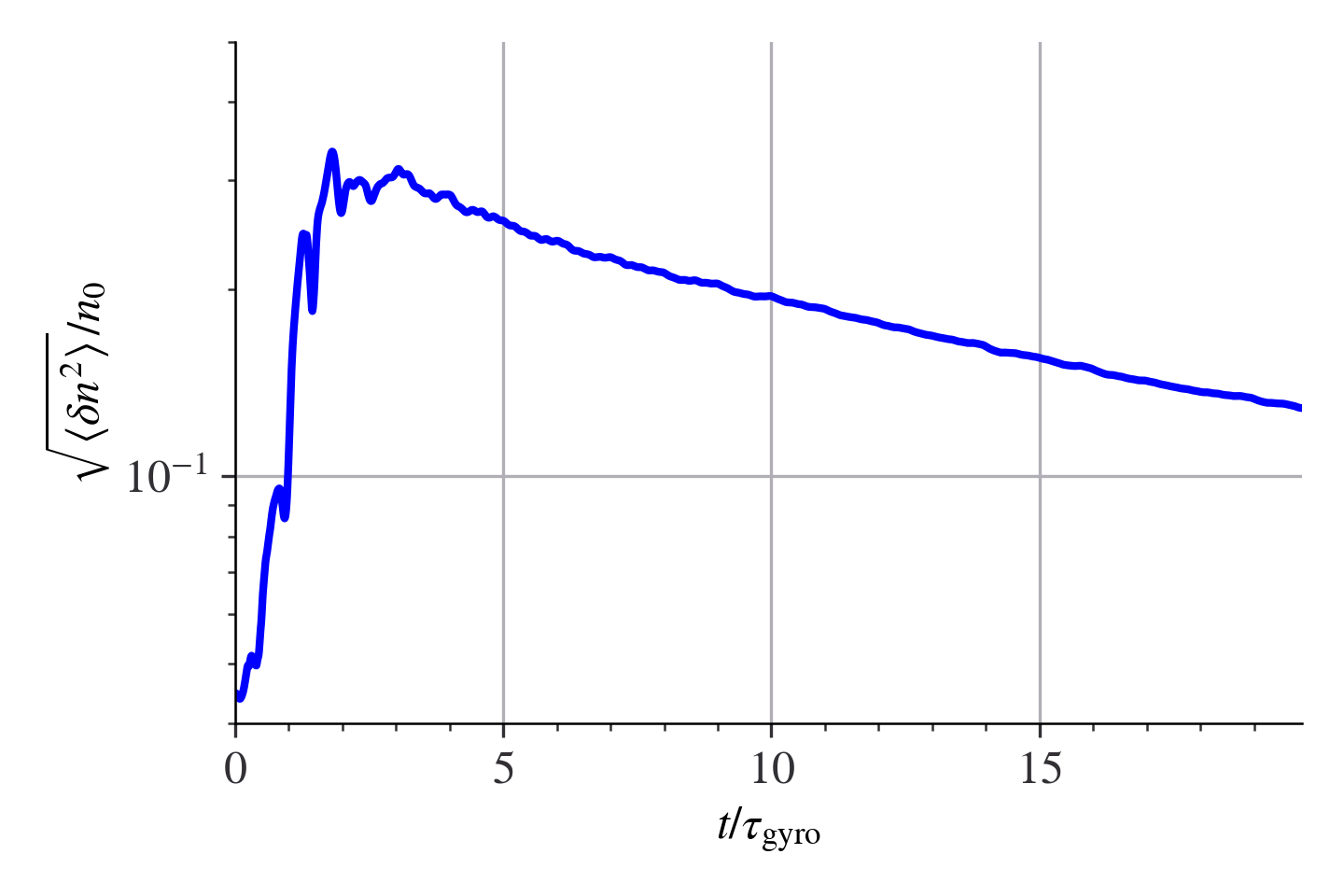}
    \caption{Normalized density variance $\sqrt{\langle \delta n^2 \rangle} / n_0$ versus time. Here, $n$ is the local density, $n_0$ is the unperturbed background density, $\delta n = n - n_0$ is the density perturbation, and $\langle \cdot \rangle$ represents spatial averaging over the simulation domain.}
    \label{fig:rms-density}
\end{figure}

Density perturbations are both macroscopically observable by spacecraft and a diagnostic of underlying wave physics, making them valuable for characterizing different wave modes. Figure \ref{fig:rms-density} shows the temporal evolution of density variance spatially averaged over the simulation domain. The density variance exhibits similar behavior to the parallel magnetic field energy $B_\parallel^2$ [Figure \ref{fig:energy-budget}(b)]: exponential growth during $0 < t / \tau_{\mathrm{gyro}} < 2$ (modulated by ion gyrophase), saturation near $t / \tau_{\mathrm{gyro}} = 2$, followed by decay on a longer timescale.

\begin{figure}[tphb]
    \centering
    \includegraphics[width=\linewidth]{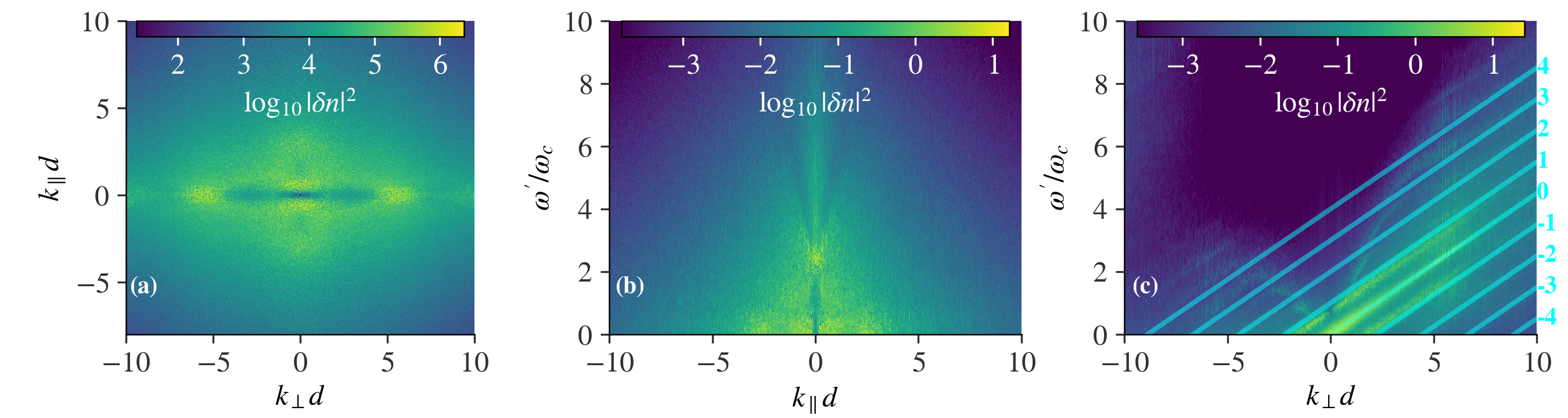}
    \caption{Fourier analysis of density perturbations. (a, b, c) The power of density perturbations in $(k_\perp, k_\parallel)$, $(k_\parallel, \omega')$, and $(k_\perp, \omega')$ space, respectively. \change{White}{Cyan} lines in Panel (c) represent ion gyrofrequency harmonics in the center-of-mass frame, $\omega / \omega_c = (\omega' - k_\perp v_{\mathrm{cm}}) / \omega_c = 0, \pm 1, \pm 2, \pm 3, \pm 4$, where $\omega$ and $\omega'$ denote the wave frequencies in the center-of-mass and \change{Moon's}{moon's} frames, respectively.}
    \label{fig:disps-density}
\end{figure}

\begin{figure}[tphb]
    \centering
    \includegraphics[width=\linewidth]{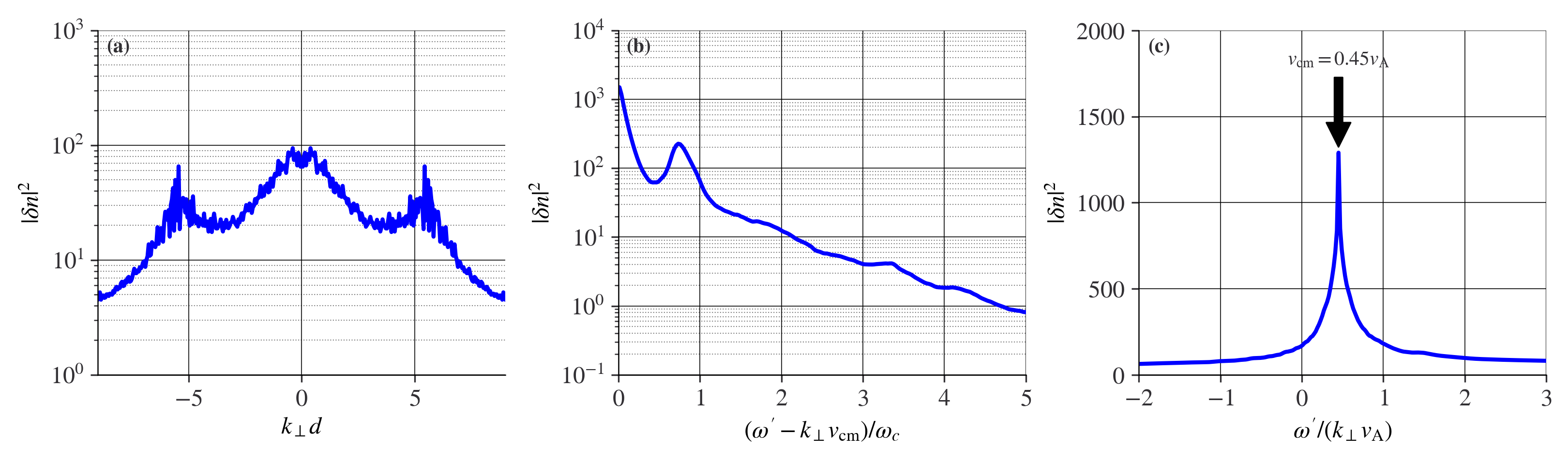}
    \caption{Characteristics of density perturbations associated with different wave modes. (a) Distribution of density perturbations in $k_\perp$. (b) Frequency spectrum of $\vert \delta n\vert^2$ in the center-of-mass frame. The main and secondary peaks are associated with the mirror mode at $\omega = \omega' - k_\perp v_{\mathrm{cm}} = 0$ and the EMIC waves at $\omega = \omega' - k_\perp v_{\mathrm{cm}} = 0.7 \omega_c$. (c) $\vert \delta n\vert^2$ as a function of perpendicular phase velocity, showing the mirror mode at $\omega'/k_\perp = v_{\mathrm{cm}} = 0.45 v_{\mathrm{A}}$. Because of the small $k_\perp$ of EMIC waves, their perpendicular phase velocities are out of range in this panel.}
    \label{fig:disps-density-lines}
\end{figure}

Following the presentation in Section \ref{sec:wave-excitation}, Figure \ref{fig:disps-density} shows the Fourier spectrum of density perturbations $\delta n$. The density perturbations exhibit a bimodal $k_\perp$ distribution: a long-wavelength population clustered at $\vert k_\perp d \vert < 2.5$ and a short-wavelength population centered near $k_\perp d = \pm 5$ [Figures \ref{fig:disps-density}(a) and \ref{fig:disps-density-lines}(a)]. Wave mode identification reveals that the long-wavelength perturbations are dominated by quasi-static mirror modes ($\omega \approx 0$) and electromagnetic ion cyclotron (EMIC) waves ($\omega / \omega_c \approx 0.7$) [Figures \ref{fig:disps-density}(c) and \ref{fig:disps-density-lines}(b)], whereas the short-wavelength perturbations are ion Bernstein waves with frequencies between successive ion gyroharmonics.

\begin{figure}[tphb]
    \centering
    \includegraphics[width=\linewidth]{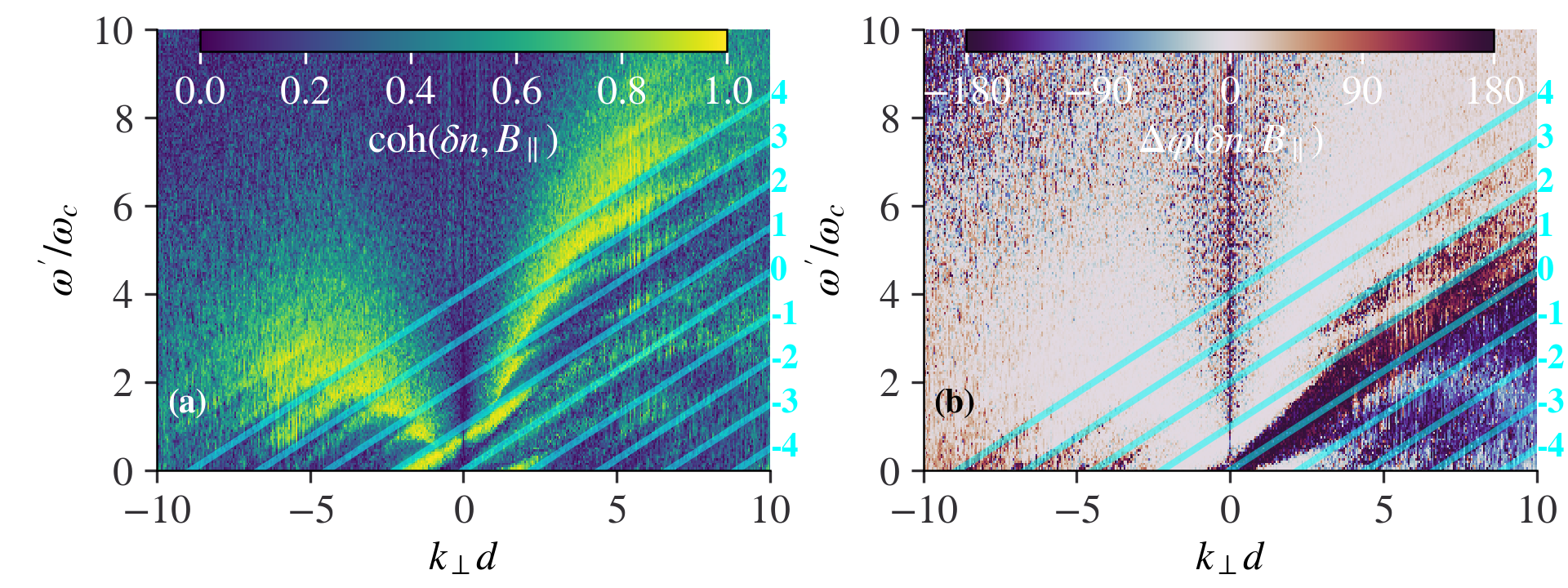}
    \caption{Spectra of (a) coherence and (b) phase difference between density and parallel magnetic field perturbations. \change{White}{Cyan} lines in both panels represent ion gyrofrequency harmonics in the center-of-mass frame, $\omega / \omega_c = (\omega' - k_\perp v_{\mathrm{cm}}) / \omega_c = 0, \pm 1, \pm 2, \pm 3, \pm 4$.}
    \label{fig:density-bpara}
\end{figure}

The dominant density perturbations are the long-wavelength mirror-mode waves, which are stationary in the plasma rest frame [Figures \ref{fig:disps-density}(c) and \ref{fig:disps-density-lines}(c)]. For a bi-Maxwellian plasma, the phase relation between density perturbations $\delta n$ and parallel magnetic field perturbations $B_\parallel$ is given by \cite{Hasegawa69}
\begin{linenomath*}
    \begin{align}
        \frac{\delta n}{n_0} \sim \left(1 - \frac{\beta_\perp}{\beta_\parallel}\right) \frac{B_\parallel}{B_0} .
    \end{align}
\end{linenomath*}
At saturation, the density perturbation amplitude $\vert \delta n / n_0 \vert \sim 0.3$ is consistent with the parallel magnetic field amplitude $\vert B_\parallel / B_0 \vert \sim 0.03$, given $\beta_\perp / \beta_\parallel \lesssim (v_{\mathrm{cr}}/2)^2 / v_{T,\mathrm{cr}}^2 = 29$ at $t = 0$. Furthermore, since $\beta_\perp > \beta_\parallel$ [Figure \ref{fig:fdists}], the theory predicts that $\delta n$ and $B_\parallel$ should be anticorrelated (out of phase). To test this prediction, we compute the coherence and phase difference spectra from the simulation data:
\begin{linenomath*}
    \begin{align}
        \mathrm{coh}(\delta n, B_\parallel) &= \frac{\vert \langle \delta n (\omega, k_\perp) \cdot B_\parallel^* (\omega, k_\perp) \rangle_{z} \vert^2}{\langle \vert \delta n (\omega, k_\perp) \vert^2 \rangle_z \cdot \langle \vert B_\parallel (\omega, k_\perp) \vert^2 \rangle_z} ,\\
        \Delta \varphi (\delta n, B_\parallel) &= \arg\left[ \langle \delta n (\omega, k_\perp) \cdot B_\parallel^* (\omega, k_\perp) \rangle_{z} \right] ,
    \end{align}
\end{linenomath*}
where $\langle \cdot \rangle_z$ denotes the ensemble average over $z$ and $\arg[\cdot]$ denotes the argument (phase angle) of a complex number. The analysis confirms the predicted anticorrelation, yielding $\Delta \varphi \approx \pm 180^\circ$ at $\omega = \omega' - k_\perp v_{\mathrm{cm}} = 0$ for mirror modes [Figure \ref{fig:density-bpara}(b)], with moderate coherence [Figure \ref{fig:density-bpara}(a)].

EMIC waves at $\omega = \omega' - k_\perp v_{\mathrm{cm}} = 0.7 \omega_c$ generate moderate density perturbations, with power density $\vert \delta n\vert^2$ approximately one order of magnitude lower than mirror modes [Figure \ref{fig:disps-density-lines}(b)]. Unlike the pressure-balance mechanism in mirror modes, EMIC density perturbations result from parallel electric fields [Figure \ref{fig:disps-elines}] that cause ion bunching along magnetic field lines, producing approximately in-phase $\delta n$ and $B_\parallel$ perturbations [Figure \ref{fig:density-bpara}].

Ion Bernstein waves exhibit the smallest density perturbations among the three wave modes [Figure \ref{fig:disps-density-lines}(b)]. At long wavelengths ($k_\perp d \lesssim 2$), the dispersion follows the fast magnetosonic relation ($\omega \propto k_\perp^2$), and $\delta n$ and $B_\parallel$ show coherent, in-phase behavior characteristic of collective fluid compression [Figure \ref{fig:density-bpara}]. At shorter wavelengths ($k_\perp d \gtrsim 3$), corresponding to the kinetic Bernstein regime with frequencies between ion gyroharmonics, the phase relationship becomes inconsistent and coherence is suppressed. This transition reflects finite Larmor radius effects: when $k_\perp \rho \gtrsim 1$, \change{gyrophase mixing among ions within a wavelength causes destructive interference, exponentially suppressing the density response}{ions distributed across a range of gyrophases within a wavelength sample the wave field with a spread of phases, and their contributions to the density perturbation partially cancel upon gyroaveraging. This effect is captured by the Bessel-function gyroaveraging factor $J_0(k_\perp \rho)$, which decays algebraically as $(k_\perp\rho)^{-1/2}$ with oscillatory sign changes for $k_\perp \rho \gtrsim 1$, rather than contributing coherently as in the long-wavelength fluid limit }\cite{bookStix62}.

\section{Field-particle correlation analysis}\label{sec:fpcorr}
\subsection{Definition}
To identify velocity space structures responsible for wave growth and their associated resonances, we perform a field-particle correlation analysis \cite{klein2016measuring,howes2017diagnosing,klein2017diagnosing}. We define the field-particle correlation function
\begin{linenomath*}
    \begin{align}
        \mathcal{C}(\mathbf{v}) =  \int \mathrm{d}t \int \mathrm{d}^3x\, f(\mathbf{x}, \mathbf{v}, t) \mathbf{v} \cdot \mathbf{E}(\mathbf{x}, t) ,
    \end{align}
\end{linenomath*}
where the temporal integration is performed over a duration sufficiently long to average out oscillations on wave timescales, and the spatial integration is conducted over the entire simulation domain. In particle-based simulations\add{, including hybrid-PIC methods in which ions are represented as discrete particles }\cite{cerri2021stochastic}, we calculate $\mathcal{C}(\mathbf{v})$ by interpolating electric fields to particle positions, computing $\mathbf{v} \cdot \mathbf{E}$ for each particle, and depositing this quantity onto velocity space grids. Regions with $\mathcal{C} < 0$ and $\mathcal{C} > 0$ represent contributions to wave growth and damping, respectively \cite{klein2025dielectric}. We integrate $\mathcal{C}(\mathbf{v})$ over gyrophase to obtain its distribution in the $(v_\perp, v_\parallel)$ plane, since only the gyrophase-averaged (DC) component of $\mathcal{C}(\mathbf{v})$ contributes to net wave growth or damping; nongyrotropic structures in $f(\mathbf{x}, \mathbf{v}, t)$ drive net wave growth only when they project onto this DC component. We use this analysis to address two questions: (1) which wave-particle resonances regulate the growth and damping of each wave mode, and (2) how ion nongyrotropy drives wave growth and simultaneously relaxes. In Section~\ref{subsec:fpcorr-simulation}, we compute $\mathcal{C}(\mathbf{v})$ directly from simulation data, while Section~\ref{subsec:fpcorr-theory} examines the contribution of ion nongyrotropy to wave growth and damping through wave-particle resonances.

\subsection{Diagnosing field-particle energy transfer in the simulation}\label{subsec:fpcorr-simulation}
\begin{figure}[tphb]
    \centering
    \includegraphics[width=\linewidth]{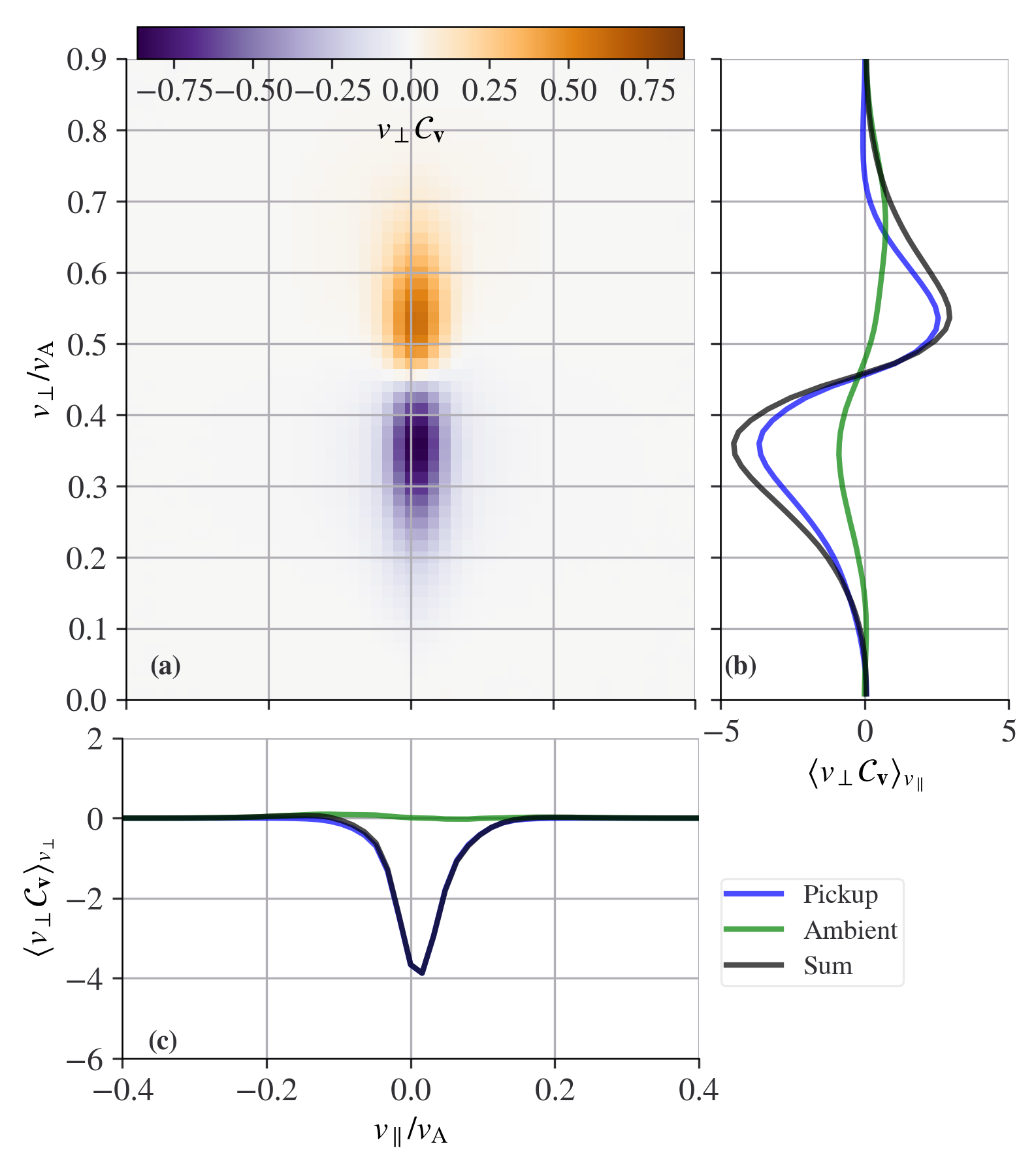}
    \caption{Field-particle correlation function $\mathcal{C}(\mathbf{v})$ evaluated before wave saturation over the interval $0 < t/\tau_{\mathrm{gyro}} < 2$. Note that $v_\perp \mathcal{C}(\mathbf{v})$ is plotted rather than $\mathcal{C}(\mathbf{v})$ itself, as the factor of $v_\perp$ accounts for the azimuthal coordinate scale factor in the cylindrical velocity space integral. (a) $v_\perp \mathcal{C}(\mathbf{v})$ in the $(v_\perp, v_\parallel)$ plane. (b)--(c) Marginal distributions obtained by integrating $v_\perp \mathcal{C}(\mathbf{v})$ over $v_\parallel$ and $v_\perp$, respectively. Blue and green lines denote contributions from pickup and ambient ions, respectively, and their sum is shown in black.}
    \label{fig:fpcorr-sat}
\end{figure}

\begin{figure}[tphb]
    \centering
    \includegraphics[width=\linewidth]{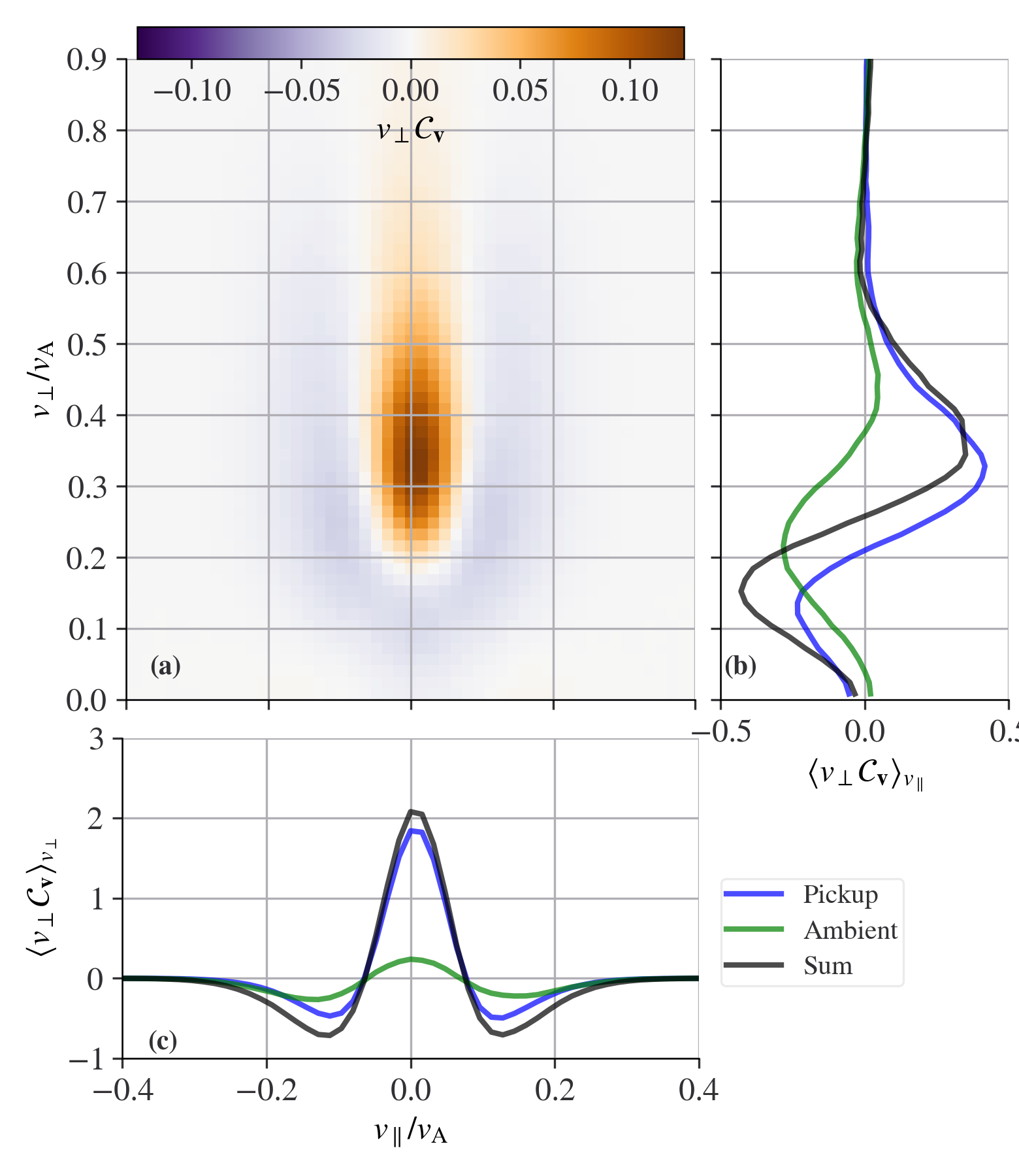}
    \caption{Field-particle correlation function $\mathcal{C}(\mathbf{v})$ after wave saturation, for the time period $t/\tau_{\mathrm{gyro}} > 2$. The format is the same as for Figure \ref{fig:fpcorr-sat}.}
    \label{fig:fpcorr-postsat}
\end{figure}

Figures \ref{fig:fpcorr-sat} and \ref{fig:fpcorr-postsat} show $\mathcal{C}(\mathbf{v})$ in the two-dimensional velocity space $(v_\perp, v_\parallel)$ for the time periods before ($0 < t/\tau_{\mathrm{gyro}} < 2$) and after ($t/\tau_{\mathrm{gyro}} > 2$) wave saturation. During $0 < t/\tau_{\mathrm{gyro}} < 2$, ions with $v_\perp / v_{\mathrm{A}} < 0.45$\remove{, below the center-of-mass velocity,} transfer energy to the waves, while ions with $v_\perp / v_{\mathrm{A}} > 0.45$ damp the waves [Figures \ref{fig:fpcorr-sat}(a) and \ref{fig:fpcorr-sat}(b)]. These correspond to velocity space structures with $\partial_{v_\perp} f > 0$ and $\partial_{v_\perp} f < 0$\add{ in the center-of-mass frame} [Figure \ref{fig:fdists}(e)], respectively. The wave growth from regions with $\partial_{v_\perp} f > 0$ exceeds the damping from regions with $\partial_{v_\perp} f < 0$ [Figure \ref{fig:fpcorr-sat}\change{(c)}{(b)}], resulting in net wave growth\add{ [Figure }\ref{fig:fpcorr-sat}\add{(c)]}. After saturation at $t/\tau_{\mathrm{gyro}} > 2$, ions with $\vert v_\parallel \vert / v_{\mathrm{A}} < 0.08$ result in net damping, while ions with $\vert v_\parallel \vert / v_{\mathrm{A}} > 0.08$ produce net wave growth due to pitch angle anisotropies [Figures \ref{fig:fpcorr-postsat}(a) and \ref{fig:fpcorr-postsat}(c)]. The damping contribution from the $\vert v_\parallel \vert / v_{\mathrm{A}} < 0.08$ region (where $\mathcal{C} > 0$) exceeds the growth contribution from the $\vert v_\parallel \vert / v_{\mathrm{A}} > 0.08$ region (where $\mathcal{C}<0$) [Figure \ref{fig:fpcorr-postsat}(c)], resulting in net wave damping during this period.

\subsection{Generalized formalism of $\mathcal{C}(\mathbf{v})$ for nongyrotropic distributions}\label{subsec:fpcorr-theory}
We decompose the total distribution function $f(\mathbf{x}, \mathbf{v}, t)$ into a spatially averaged component $g(\mathbf{v}, t)$ and a rapidly fluctuating component $\delta f(\mathbf{x}, \mathbf{v}, t)$ following quasi-linear theory, where only the fluctuating component contributes to $\mathcal{C}(\mathbf{v})$ (see \ref{sec:fpcorr-ql}). The spatially averaged component $g(\mathbf{v}, t)$ is nongyrotropic and is expanded in gyrophase harmonics (\ref{sec:avg-gdist}):
\begin{linenomath*}
    \begin{align}
        g(\mathbf{v}, t) = \sum_l g_l(v_\perp, v_\parallel, t) e^{i l (\phi + \omega_c t)} ,
    \end{align}
\end{linenomath*}
where $l$ denotes the gyrophase harmonic index. The rapidly fluctuating component is Fourier-analyzed in space and obtained by integrating along unperturbed particle orbits (\ref{sec:pert-df}):
\begin{linenomath*}
    \begin{align}\label{eq:deltaf-nongyro-main}
        \delta f_{\mathbf{k}} =& -\frac{i Q}{M} \sum_l \exp\left[- i (\nu_{\mathbf{k}} - l \omega_c) t + i l \phi \right] \sum_{m,n} \frac{J_m e^{i(m-n)(\phi - \psi)}}{\lambda_{\mathbf{k}} - n \omega_c} \nonumber \\
        \cdot& \left[\frac{E_\mathbf{k}^r e^{i \psi} J_{n+1} + E_{\mathbf{k}}^l e^{-i \psi} J_{n-1}}{\sqrt{2}} \hat{G}_\mathbf{k} + E_\mathbf{k}^\parallel J_{n} \left(\partial_{v_\parallel} + \frac{n \omega_c}{\nu_\mathbf{k} v_\perp} \hat{H}\right) \right. \nonumber \\
        &\left. - \frac{E_\mathbf{k}^r e^{i \psi} J_{n+1} - E_{\mathbf{k}}^l e^{-i \psi} J_{n-1}}{\sqrt{2}} \frac{l \lambda_\mathbf{k}}{\nu_\mathbf{k} v_\perp} + E_\mathbf{k}^\parallel J'_{n} \frac{l k_\perp v_\parallel}{\nu_\mathbf{k} v_\perp} +\frac{E_\mathbf{k}^r e^{i \psi} - E_\mathbf{k}^l e^{-i \psi}}{\sqrt{2}} J_{n} \frac{l k_\perp}{\nu_\mathbf{k}} \right] g_l ,
    \end{align}
\end{linenomath*}
where $Q$ and $M$ are the ion charge and mass, $\nu_{\mathbf{k}} = \omega_{\mathbf{k}} + i \gamma_{\mathbf{k}}$ is the complex wave frequency at wavenumber $\mathbf{k}$, $\lambda_{\mathbf{k}} = \nu_{\mathbf{k}} - k_\parallel v_\parallel$ is the Doppler-shifted frequency, $\psi$ is the azimuthal angle of $\mathbf{k}_\perp$ ($k_x = k_\perp\cos\psi$, $k_y = k_\perp\sin\psi$), $J_m$ is the $m$-th order Bessel function of argument $k_\perp v_{\perp} / \omega_c$, and $\hat{G}_{\mathbf{k}} = \partial_{v_\perp} - \frac{k_\parallel}{\nu_{\mathbf{k}}} \hat{H}$ and $\hat{H} = v_\parallel \partial_{v_\perp} - v_\perp \partial_{v_\parallel}$ are velocity-space differential operators. For a nongyrotropic distribution ($l \neq 0$), a wave electric field oscillating as $e^{-i \nu_{\mathbf{k}}t}$ drives perturbations in $\delta f_{\mathbf{k}}$ with time dependence $e^{- i (\nu_{\mathbf{k}} - l \omega_c) t}$, consistent with earlier analyses of parallel-propagating waves driven by nongyrotropic distributions \cite{sudan1965growing, brinca1993on}. The terms proportional to $l g_l$ represent the nongyrotropic contribution to wave growth or damping, and $\delta f_{\mathbf{k}}$ reduces to the gyrotropic result of \citeA{Kennel&Engelmann66} when $l = 0$. Crucially, the resonance condition is unaffected by the gyrophase harmonic index: the resonance order $n$ and the harmonic index $l$ enter independently. We define the resonance factor
\begin{linenomath*}
    \begin{align}
        \mathcal{R}(n) = -\frac{1}{\lambda_{\mathbf{k}} - n \omega_c} = -\frac{1}{\omega_{\mathbf{k}} + i \gamma_{\mathbf{k}} - k_\parallel v_\parallel - n \omega_c} .
    \end{align}
\end{linenomath*}
The field-particle correlation function can then be expressed in terms of the wave electric fields and the perturbed distribution function (\ref{sec:fpcorr-final-formalism}):
\begin{linenomath*}
    \begin{align}\label{eq:fpcorr-final-ql}
        \mathcal{C}(\mathbf{v}) =& \frac{2 \pi i Q}{M} \sum_{l,n} \int \frac{\mathrm{d}^3 k}{(2 \pi)^3} \underbrace{\mathcal{R}(n)}_{\mathrm{resonance\,\, factor}}  e^{i l \psi} \nonumber \\
        &\cdot \underbrace{ \left[ \frac{v_\perp (E_{\mathbf{k}}^l(\nu_\mathbf{k} - l \omega_c) e^{-i \psi})^* J_{n-l-1} e^{i \psi}}{\sqrt{2}} + \frac{v_\perp (E_{\mathbf{k}}^r(\nu_\mathbf{k} - l \omega_c) e^{i \psi})^* J_{n-l+1} e^{-i \psi}}{\sqrt{2}} + v_\parallel (E_{\mathbf{k}}^{\parallel}(\nu_\mathbf{k} - l \omega_c))^* J_{n-l} \right] }_{\mathbf{v} \cdot \mathbf{E}^*_{\mathbf{k}}} \nonumber \\
        &\cdot \left[\frac{E_\mathbf{k}^r(\nu_\mathbf{k}) e^{i \psi} J_{n+1} + E_{\mathbf{k}}^l(\nu_\mathbf{k}) e^{-i \psi} J_{n-1}}{\sqrt{2}} \hat{G}_\mathbf{k} + E_\mathbf{k}^\parallel(\nu_\mathbf{k}) J_{n} \left(\partial_{v_\parallel} + \frac{n \omega_c}{\nu_\mathbf{k} v_\perp} \hat{H}\right) \right. \nonumber \\
        &\underbrace{ \left. - \frac{E_\mathbf{k}^r(\nu_\mathbf{k}) e^{i \psi} J_{n+1} - E_{\mathbf{k}}^l(\nu_\mathbf{k}) e^{-i \psi} J_{n-1}}{\sqrt{2}} \frac{l \lambda_\mathbf{k}}{\nu_\mathbf{k} v_\perp} + E_\mathbf{k}^\parallel(\nu_\mathbf{k}) J'_{n} \frac{l k_\perp v_\parallel}{\nu_\mathbf{k} v_\perp} +\frac{E_\mathbf{k}^r(\nu_\mathbf{k}) e^{i \psi} - E_\mathbf{k}^l(\nu_\mathbf{k}) e^{-i \psi}}{\sqrt{2}} J_{n} \frac{l k_\perp}{\nu_\mathbf{k}}\right] g_l }_{\delta f_{\mathbf{k}}} \nonumber \\
        \equiv& \sum_{l,n} \sum_{\sigma, \sigma'} \int \frac{\mathrm{d}^3 k}{(2 \pi)^3} \mathcal{K}_{\sigma \sigma'}^{(n, l)} E_{\mathbf{k}}^\sigma (\nu_{\mathbf{k}}) E_{\mathbf{k}}^{\sigma' *} (\nu_{\mathbf{k}} - l \omega_c),
    \end{align}
\end{linenomath*}
where $\sigma, \sigma' \in \{l, r, \parallel\}$ and $\mathcal{K}_{\sigma \sigma'}^{(n, l)}$ is the field-particle correlation coefficient of resonance order $n$ and gyrophase harmonic $l$ coupling polarization $\sigma$ at frequency $\nu_{\mathbf{k}}$ to polarization $\sigma'$ at frequency $\nu_{\mathbf{k}} - l\omega_c$. The detailed expressions for each component of $\mathcal{K}_{\sigma \sigma'}^{(n, l)}$ are given in \ref{sec:fpcorr-final-formalism}. For a nongyrotropic distribution ($l \neq 0$), the electric field components $E_{\mathbf{k}}^{l,r,\parallel}(\nu_{\mathbf{k}})$ at frequency $\nu_{\mathbf{k}}$ couple to $E_{\mathbf{k}}^{l,r,\parallel}(\nu_{\mathbf{k}} - l \omega_c)$ at frequency $\nu_{\mathbf{k}} - l \omega_c$, opening additional channels for net field-particle energy transfer that are absent in the gyrotropic case ($l = 0$).

\begin{figure}[tphb]
    \centering
    \includegraphics[width=\linewidth]{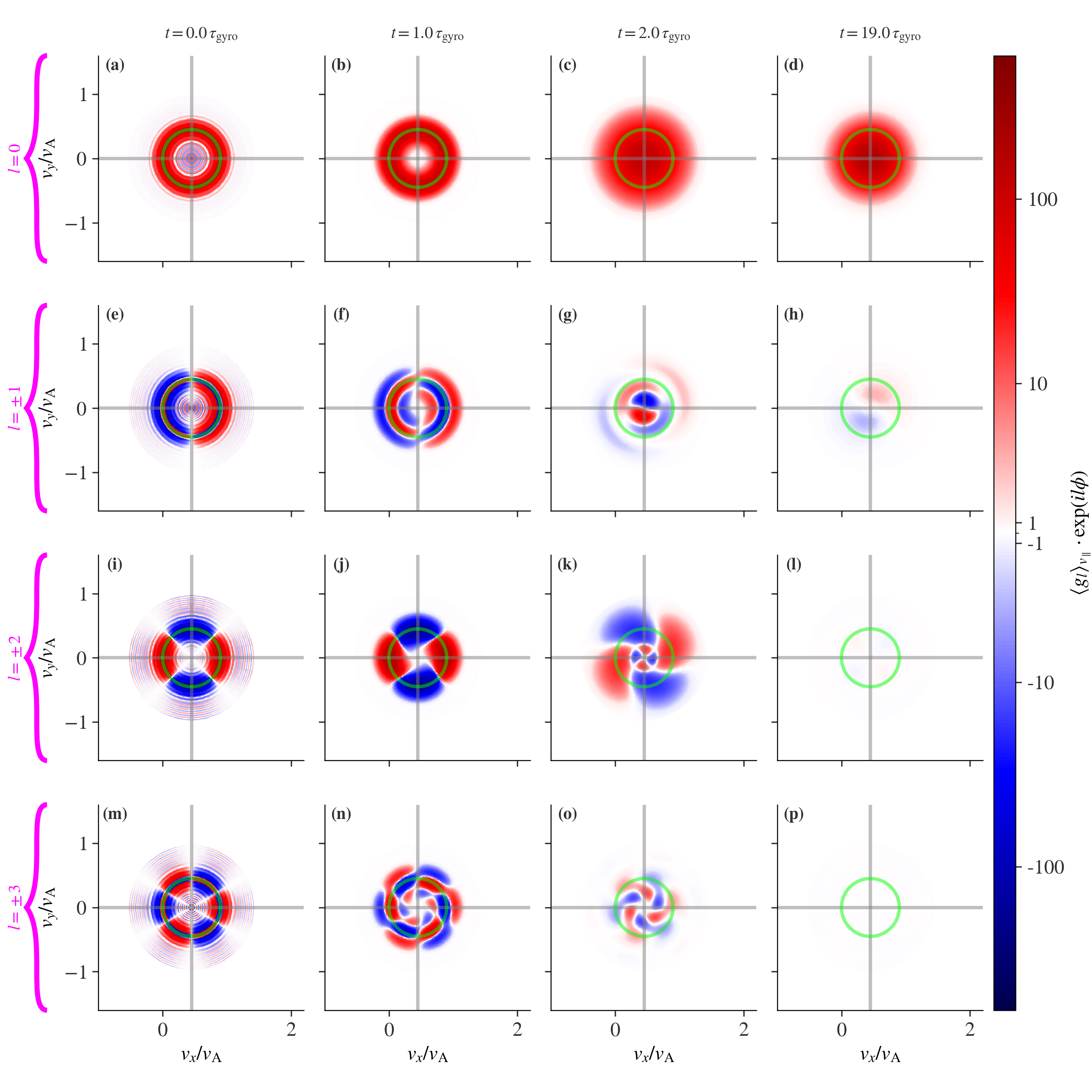}
    \caption{Decomposition of the spatially averaged distribution $g(\mathbf{v}, t)$ into gyrophase harmonics $g_l e^{i l \phi}$ at four representative times. Columns from left to right correspond to $t/\tau_{\mathrm{gyro}} = 0, 1, 2, 19$. Rows from top to bottom show the fundamental mode ($l=0$) and higher harmonics ($l = \pm 1, \pm 2, \pm 3$). Distributions are shown in the perpendicular velocity plane, integrated over $v_\parallel$. The \add{intersection of the }two straight \change{green}{gray} lines denote the center-of-mass velocity. The green circle marks the unperturbed gyro-orbit of pickup ions with initial velocity $v_x = v_y = 0$.}
    \label{fig:f-bessel-fourier}
\end{figure}

Figure \ref{fig:f-bessel-fourier} shows the first few harmonics of $g_l$ at four representative times in our simulation. To obtain these harmonics, we employ Fourier decomposition in $\phi$ combined with Bessel decomposition in $v_\perp$: $g(\mathbf{v}, t) = \sum_l g_l(v_\perp, v_\parallel, t) e^{i l (\phi + \omega_c t)} = \sum_{l, n} a_{l,n}(v_\parallel, t) J_{n}(\frac{\alpha_{l,n}}{v_{\max}} v_\perp) e^{i l (\phi + \omega_c t)}$, where $a_{l,n}(v_\parallel, t)$ is the coefficient, $\alpha_{l,n}$ is the $n$-th positive zero of $J_l$, and $v_{\max}$ is the upper bound of $v_\perp$. At $t = 0$, the fundamental mode exhibits a ring distribution symmetric about the center of mass and peaked at $v_\perp = 0.45 v_{\mathrm{A}}$\add{ [Figure }\ref{fig:f-bessel-fourier}\add{(a)]}, \change{with significant higher harmonics present}{while higher harmonics ($l>0$) show segmented rings} due to nongyrotropy [Figures \ref{fig:f-bessel-fourier}(e), \ref{fig:f-bessel-fourier}(i), and \ref{fig:f-bessel-fourier}(m)]. During wave growth ($t / \tau_{\mathrm{gyro}} = 1$) and upon wave saturation ($t / \tau_{\mathrm{gyro}} = 2$), the ring distribution $g_0$ is scattered toward a Maxwellian distribution, and the positive gradient $\partial_{v_{\perp}} g_0 > 0$ diminishes [Figures \ref{fig:f-bessel-fourier}(b) and \ref{fig:f-bessel-fourier}(c)]. For higher harmonics $g_l e^{i l \phi}$, ions at different $v_\perp$ exhibit different gyrophase evolution over a fixed time interval (i.e., gyrophase desynchronization) [Figures \ref{fig:f-bessel-fourier}(f-g), \ref{fig:f-bessel-fourier}(j-k), and \ref{fig:f-bessel-fourier}(n-o)]. This velocity-dependent desynchronization leads to phase mixing, causing the higher harmonics to diminish by the end of the simulation [Figures \ref{fig:f-bessel-fourier}(h), \ref{fig:f-bessel-fourier}(l), and \ref{fig:f-bessel-fourier}(p)].

\subsection{Velocity-space structures driving EMIC, ion Bernstein, and mirror-mode instabilities}
\begin{figure}[tphb]
    \centering
    \includegraphics[width=\linewidth]{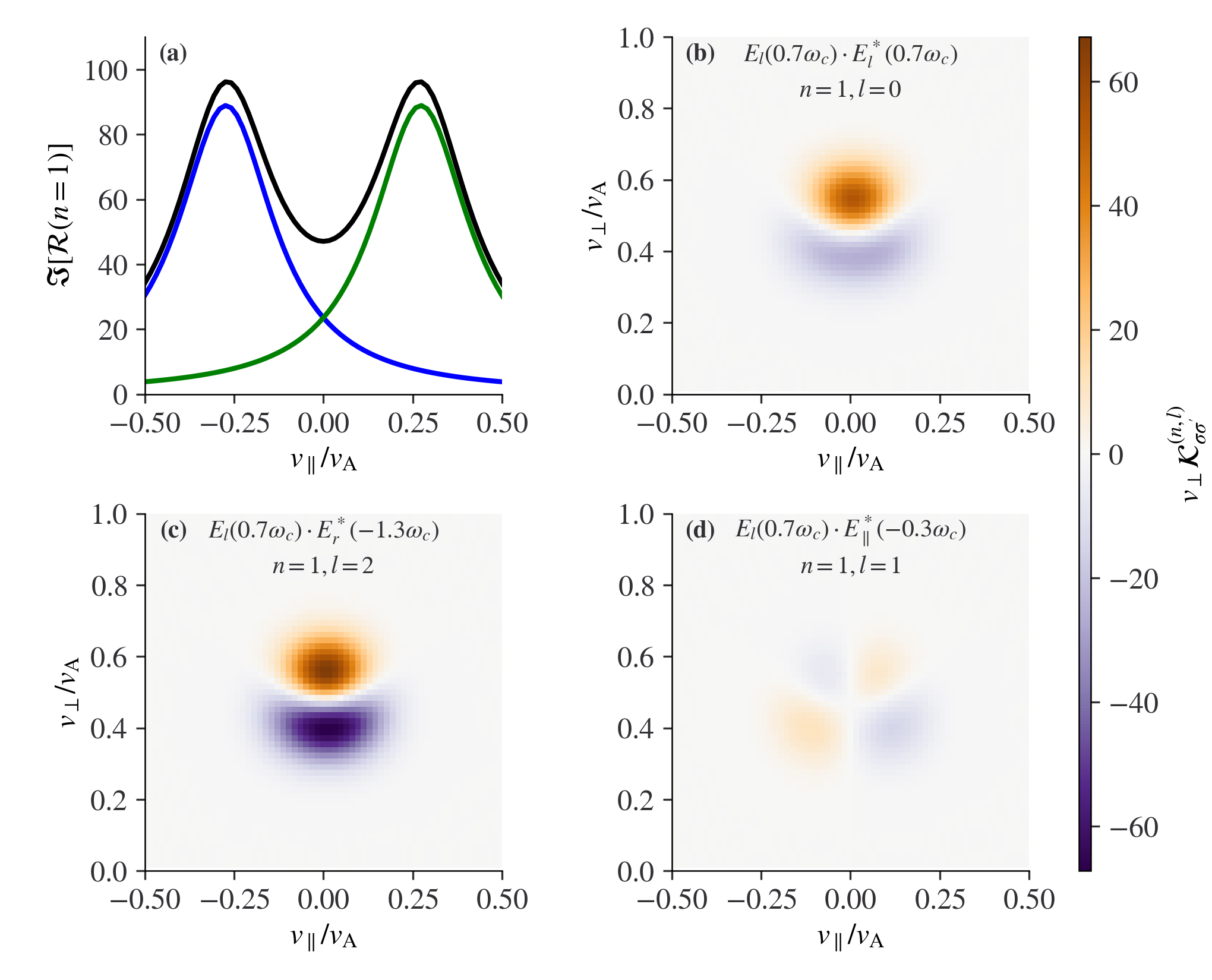}
    \caption{Resonance factor and coupling coefficients of the field-particle correlation function for EMIC waves. (a) Imaginary part of the resonance factor $\mathcal{R}(n=1) = -1 / (\omega - k_\parallel v_\parallel - \omega_c)$ for the fastest-growing EMIC mode ($k_\parallel d = \pm 1.1$, $k_\perp = 0$, $\omega / \omega_c = 0.7$, $\gamma / \omega_c = 0.18$), for $k_\parallel d = 1.1$ (blue), $k_\parallel d = -1.1$ (green), and their sum (black). (b)--(d) Coupling coefficients $\mathcal{K}_{ll}^{(1,0)}$, $\mathcal{K}_{lr}^{(1,2)}$, and $\mathcal{K}_{l\parallel}^{(1,1)}$ [Equation~\eqref{eq:fpcorr-final-ql}] for the same mode. As in Figures~\ref{fig:fpcorr-sat} and \ref{fig:fpcorr-postsat}, each coefficient is multiplied by the azimuthal scale factor $v_\perp$ so that contributions at different $v_\perp$ are weighted equally in the velocity-space integration.}
    \label{fig:fpcorr-emic}
\end{figure}

We are now in a position to use the field-particle correlation formalism [Equation \eqref{eq:fpcorr-final-ql}] and the Fourier decomposition $g_l$ [Figure \ref{fig:f-bessel-fourier}] to understand how velocity-space structures and resonances control the growth and damping of different wave modes. We first focus on the EMIC waves. Based on Section \ref{sec:wave-excitation}, the dominant EMIC mode has $k_\parallel d = \pm 1.1$, $k_\perp = 0$, and $\omega / \omega_c = 0.7$. Based on the temporal evolution of $B_\perp^2$ [Figure \ref{fig:energy-budget}(b)], this mode is amplified by two orders of magnitude in two gyroperiods before saturation, giving a growth rate of approximately $\gamma / \omega_c = \ln[B_{\perp}^2 (t = 2 \tau_{\mathrm{gyro}}) / B_{\perp}^2 (t = 0)] / 8 \pi = 0.18$. The most viable resonance for EMIC wave excitation is the normal cyclotron resonance with $n=1$. Figure \ref{fig:fpcorr-emic}(a) shows the imaginary part of $\mathcal{R}(n=1)$ (which contributes to the real part of $\mathcal{C}(\mathbf{v})$) for this EMIC resonance as a function of $v_\parallel$. The resonance velocity is $v_r  = (\omega_c - \omega) / k_\parallel = \pm 0.27 v_{\mathrm{A}}$. Because $\vert v_r \vert$ is much larger than the thermal velocities of ambient and pickup ions ($v_{T,\mathrm{cr}} = 0.083 v_{\mathrm{A}}$ and $v_{T,\mathrm{pu}} = 0.018 v_{\mathrm{A}}$), the small number of ions at $v_r$ cannot make significant contributions to wave growth or damping. On the other hand, the spread of $\Im[\mathcal{R}(1)]$ in $v_\parallel$ is $\gamma / k_\parallel = 0.16 v_{\mathrm{A}}$, allowing the bulk of the ion distribution near $v_\parallel = 0$ to contribute considerably.

For parallel-propagating EMIC waves, only three nonzero coefficients exist in the field-particle correlation function [Equation \eqref{eq:fpcorr-final-ql}]: $\mathcal{K}_{ll}^{(1, 0)}$, with resonance order $n=1$ and gyrophase harmonic $l=0$, arising from the self-coupling $E_l(0.7 \omega_c) \cdot E_l^*(0.7 \omega_c)$ [Figure \ref{fig:fpcorr-emic}(b)]; $\mathcal{K}_{lr}^{(1,2)}$, with $n=1$ and $l=2$, coupling $E_{l}(0.7 \omega_c)$ to $E^*_r(-1.3\omega_c) = E^*_l(1.3\omega_c)$ [Figure \ref{fig:fpcorr-emic}(c)]; and $\mathcal{K}_{l\parallel}^{(1,1)}$, with $n=1$ and $l=1$, coupling $E_{l}(0.7 \omega_c)$ to $E^*_\parallel(-0.3\omega_c) = E^*_\parallel(0.3\omega_c)$ [Figure \ref{fig:fpcorr-emic}(d)]. For $\vert v_\parallel \vert \ll v_{T,\mathrm{cr}} = 0.083 v_{\mathrm{A}}$, the sign of each coupling coefficient is governed by the perpendicular velocity gradient $\partial_{v_\perp} g_l$, whereas for $\vert v_\parallel \vert \gtrsim v_{T,\mathrm{cr}}$, the shape of the coefficients is instead controlled by the pitch-angle gradient operator $\hat{H} = v_\parallel \partial_{v_\perp} - v_\perp \partial_{v_\parallel}$ [Equation \eqref{eq:deltaf-nongyro-main}].

We find that the gyrotropic part of the distribution ($g_0$) has a net damping effect on EMIC waves [see Figure \ref{fig:fpcorr-emic}(b) and $\int \mathrm{d} v_\parallel \int \mathrm{d} v_\perp \, v_\perp \mathcal{K}_{ll}^{(1, 0)} > 0$], whereas the growth of EMIC waves is predominantly driven by the second gyrophase harmonic $g_2$ through the cross-coupling $E_{l}(0.7 \omega_c) \cdot E^*_r(-1.3\omega_c)$ [Figure \ref{fig:fpcorr-emic}(c)]. The contribution of the first gyrophase harmonic $g_1$ through the coupling $E_{l}(0.7 \omega_c) \cdot E^*_\parallel(-0.3\omega_c)$ is smaller compared to $g_2$ and $g_0$ [Figure \ref{fig:fpcorr-emic}(d)]. This result demonstrates that nongyrotropy is essential for EMIC wave growth in this system: the gyrotropic distribution alone would produce net damping, and it is specifically the gyrophase asymmetry encoded in $g_2$ that enables energy transfer from ions to waves through coupling between different wave polarizations and frequencies. Without this nongyrotropic structure, the observed rapid wave amplification would not occur.

\begin{figure}[tphb]
    \centering
    \includegraphics[width=\linewidth]{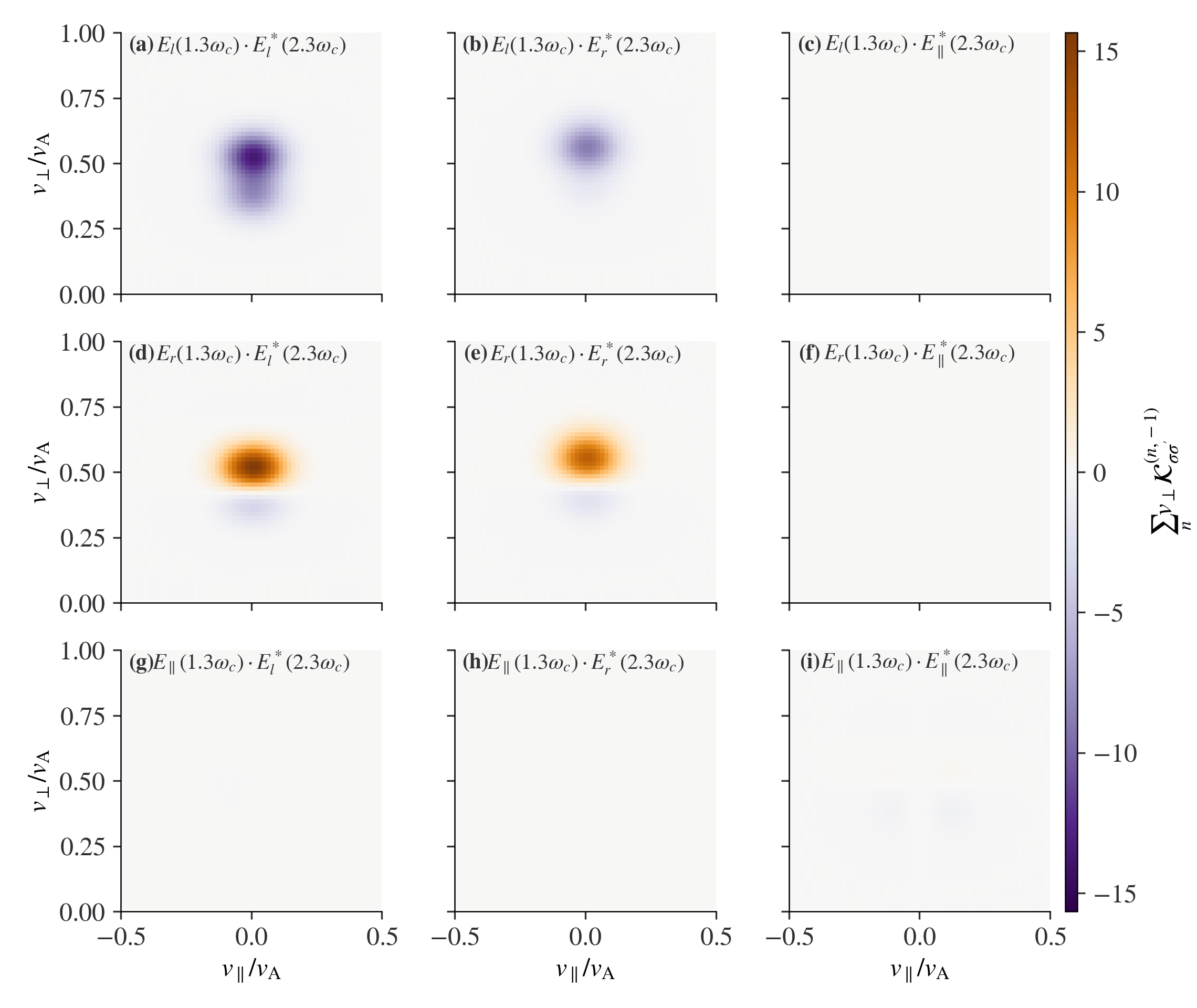}
    \caption{Coupling coefficients of the field-particle correlation function for ion Bernstein waves. The nine panels display the coupling coefficients $\sum_n v_\perp\mathcal{K}_{\sigma \sigma'}^{(n, -1)}$ for the fastest growing ion Bernstein mode with gyrophase harmonic distribution $g_{-1}$, where the rows correspond to $\sigma = l, r, \parallel$ and columns correspond to $\sigma' = l, r, \parallel$ (left to right: left-handed, right-handed, parallel components). The Jacobian factor $v_\perp$ is multiplied to each coupling coefficient, and the summation is over resonance orders from $n=-7$ to $n=7$.}
    \label{fig:fpcorr-ibw-lminus1}
\end{figure}

We examine the velocity space signatures responsible for the growth of ion Bernstein waves. The fastest growing ion Bernstein mode has $k_\parallel = 0$, $k_\perp d = 5.3$, and $\omega / \omega_c = 1.3$, with a growth rate of approximately $\gamma / \omega_c = 0.18$ prior to saturation, based on the temporal evolution of $B_\parallel^2$ [Figure~\ref{fig:energy-budget}(b)]. For perpendicularly propagating waves ($k_\parallel = 0$), the resonance factor is constant for each resonance order $n$, i.e., $\mathcal{R}(n) = -1/(\omega + i\gamma - n \omega_c) = \mathrm{constant}$, meaning that ion Bernstein waves are driven by ions throughout the entire velocity distribution \cite{chen2015wave}. We therefore sum the field-particle correlation coefficient over resonance order: $\sum_{n=-7}^{7} \mathcal{K}_{\sigma \sigma'}^{(n, l)}$ for each gyrophase harmonic distribution $g_l$, where the summation limits are chosen to ensure convergence. Figure~\ref{fig:fpcorr-ibw-lminus1} shows the nine components of $\sum_{n=-7}^{7} v_\perp \mathcal{K}_{\sigma \sigma'}^{(n, -1)}$ ($\sigma, \sigma' \in \{l, r, \parallel\}$) for the gyrophase harmonic $g_{-1}(v_\perp, v_\parallel, t=0)$, through which the mode at $\omega / \omega_c = 1.3$ couples to $\omega / \omega_c = 2.3$; additional contributions from the gyrotropic component $g_0$ and the second harmonic $g_{-2}$ are shown in Figures~\ref{fig:fpcorr-ibw-l0} and \ref{fig:fpcorr-ibw-lminus2} in \ref{sec:fpcorr-ibw-mmw}, respectively. Unlike EMIC waves with finite $k_\parallel$, for which $\mathcal{K}_{\sigma \sigma'}^{(n, l)}$ receives contributions from phase space density gradients in pitch angle, the $k_\parallel = 0$ ion Bernstein mode couples only through gradients in perpendicular velocity ($\partial_{v_\perp} g_l$) and nongyrotropy ($l g_l$). Finally, $E_\parallel$ is Landau damped [Figure~\ref{fig:fpcorr-ibw-l0}(i)], consistent with the weak $E_\parallel$ power spectral density seen in Figure~\ref{fig:disps-elines}.

\begin{figure}[tphb]
    \centering
    \includegraphics[width=\linewidth]{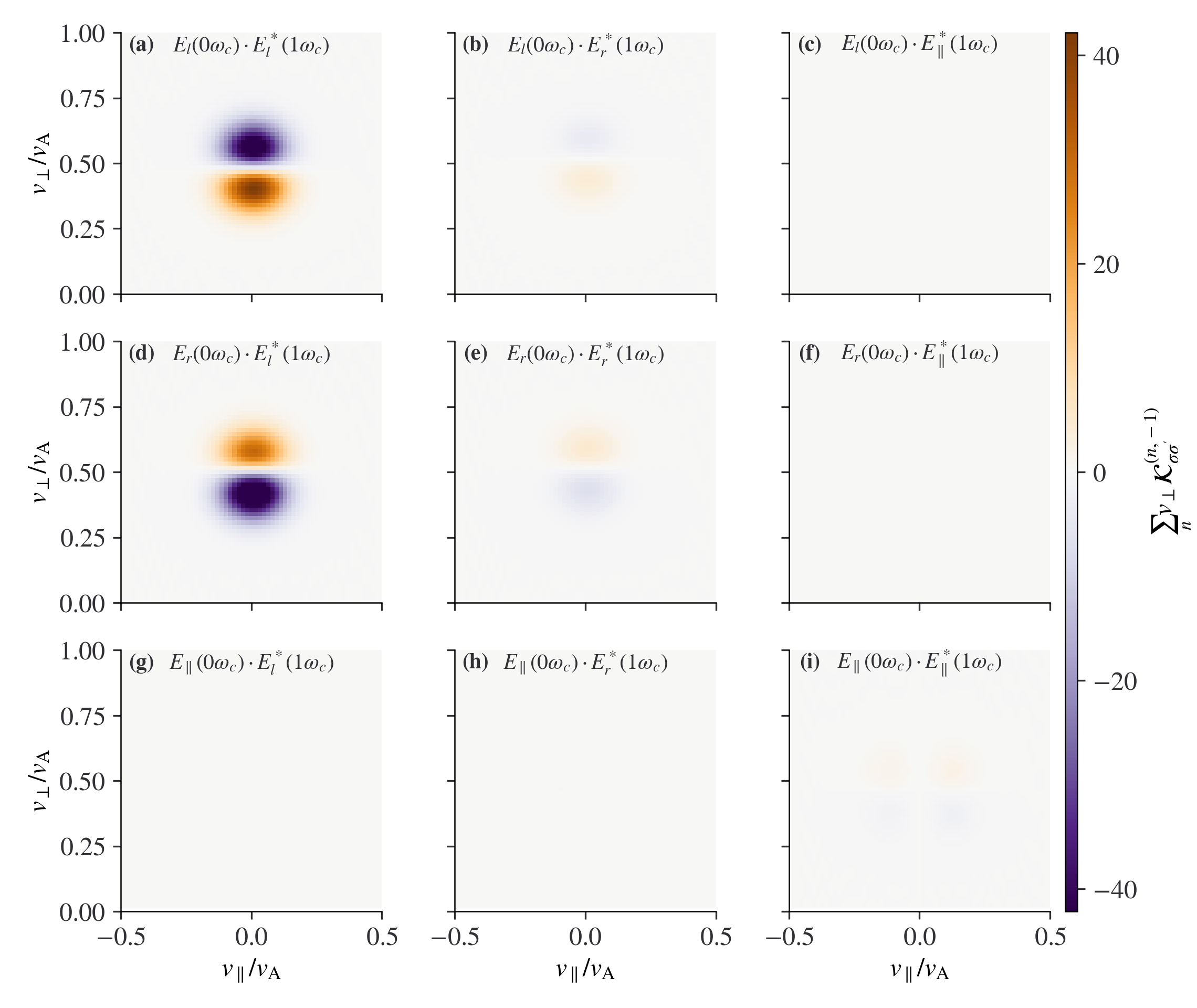}
    \caption{Coupling coefficients of the field-particle correlation function for the fastest growing mirror mode with gyrophase harmonic distribution $g_{-1}$. Panel layout and notation are the same as in Figure \ref{fig:fpcorr-ibw-lminus1}.}
    \label{fig:fpcorr-mmw-lminus1}
\end{figure}

Similarly, the fastest growing mirror mode has $k_\parallel = 0$, $k_\perp d = 1.6$, $\omega = 0$, and $\gamma / \omega_c = 0.18$. As with the ion Bernstein mode, we sum the field-particle correlation coefficients over resonance order $\sum_{n=-7}^{7} \mathcal{K}_{\sigma \sigma'}^{(n, l)}$ for each gyrophase harmonic $g_l$; the result for $g_{-1}(v_\perp, v_\parallel, t=0)$ is shown in Figure~\ref{fig:fpcorr-mmw-lminus1}, with additional contributions from the gyrotropic component $g_0$ and the second harmonic $g_{-2}$ shown in Figures~\ref{fig:fpcorr-mmw-l0} and \ref{fig:fpcorr-mmw-lminus2}, respectively. Consistent with the mirror mode polarization, the transverse components $E_l$ and $E_r$ at $\omega = 0$ receive net energy transfer from ions, while $E_\parallel$ at $\omega = 0$ is strongly Landau damped [Figure~\ref{fig:fpcorr-mmw-l0}(i)].

We find that our formalism $\mathcal{K}_{\sigma \sigma'}^{(n,l)}(\mathbf{v})$ captures key features of the structure of $\mathcal{C}(\mathbf{v})$ from the simulation data [Figures \ref{fig:fpcorr-sat} and \ref{fig:fpcorr-postsat}]: (1) The perpendicular gradients $\partial_{v_\perp} g_l$ largely control the energy transfer around $v_\parallel = 0$, while the pitch angle gradients $\hat{H} g_l$ control the energy transfer around the thermal velocity $v_\parallel = 0.1 v_{\mathrm{A}}$. (2) The nongyrotropic part of the distribution $g_l$ ($l \neq 0$) contributes significantly to the total field-particle energy transfer.
 

\section{Summary}\label{sec:summary}
In summary, using a combination of computational and theoretical approaches, we study the excitation of plasma waves and the relaxation of nongyrotropic distributions during the ion pickup process at outer planet moons. The main results are as follows.
\begin{enumerate}
    \item The ambient and pickup ions, initially separated by a corotating velocity perpendicular to the background magnetic field, naturally form a nongyrotropic distribution that provides free energy for wave growth.
    \item Both transverse and compressional magnetic perturbations are excited by the nongyrotropic distribution. These perturbations mediate the conversion from bulk kinetic to thermal energy, rapidly isotropizing the velocity distribution and completing the incorporation of the new ions into the ambient plasma within a few ion gyroperiods. Contrary to conventional assumptions, the scattering rates in perpendicular and parallel velocities can be comparable to those in the gyrophase.
    \item We derive a necessary but not sufficient condition for wave excitation: $2 M_{\mathrm{A}}^2 (1 - \eta_{\mathrm{cr}}) / (3 \beta_{\mathrm{cr}}) > 1$, where $M_{\mathrm{A}}^2 (1 - \eta_{\mathrm{cr}})$ represents the bulk kinetic energy provided by the relative motion between ambient and pickup ions and $\beta_{\mathrm{cr}}$ represents the thermal energy of the corotating ions.
    \item The transverse magnetic perturbations are predominantly quasi-parallel propagating EMIC waves, while the compressional magnetic perturbations are predominantly quasi-perpendicular propagating ion Bernstein and mirror-mode waves.
    \item Most density perturbations are caused by mirror-mode waves and are out of phase with the parallel magnetic perturbations. EMIC waves cause moderate density perturbations through parallel electric fields, while ion Bernstein waves produce the least density perturbations.
    \item We develop a generalized formalism of the field-particle correlation function for nongyrotropic distributions. This formalism demonstrates that each gyrophase harmonic $g_l (v_\perp, v_\parallel) e^{i l (\phi + \omega_c t)}$ does not alter the resonance condition, but couples the mode at frequency $\nu_{\mathbf{k}}$ to another at $\nu_{\mathbf{k}} - l \omega_c$, thereby opening additional channels for net field-particle energy transfer that are absent in the gyrotropic case ($l = 0$).
    \item Using the generalized formalism, we clarify which resonances and velocity-space structures regulate the growth and damping of different wave modes. For quasi-parallel propagating EMIC waves, the normal cyclotron resonance ($n=1$) dominates, and the gyrophase harmonic $g_2$ makes the leading contribution to wave growth. For quasi-perpendicular propagating ion Bernstein and mirror modes, higher-order resonances and gyrophase harmonics must be taken into account to obtain a complete picture of wave-particle energy transfer.
\end{enumerate}

These results demonstrate how nongyrotropic distributions regulate the excitation of different wave modes and how these waves thermalize ions in velocity space and complete the pickup process. A natural extension of this work is to establish a quantitative instability criterion by performing a large set of kinetic simulations that systematically scan the parameter space of $M_{\mathrm{A}}$, $\eta_{\mathrm{cr}}$, and $\beta_{\mathrm{cr}}$. Such a criterion would sharpen the necessary condition for wave excitation derived here [$2 M_{\mathrm{A}}^2 (1 - \eta_{\mathrm{cr}}) / (3 \beta_{\mathrm{cr}}) > 1$] and forge a connection to the well-known instability criteria for anisotropy-driven mirror and ion cyclotron instabilities \cite{Hellinger06,Bale09,shoji2009mirror}.

On the theoretical side, a linear kinetic dispersion relation can be derived for nongyrotropic distributions. Starting from the perturbed distribution function [Equation \eqref{eq:deltaf-nongyro-main}], one can calculate the perturbed current density, construct the dielectric tensor, and thereby obtain a fully general linear kinetic dispersion relation for nongyrotropic distributions. This would generalize the special cases previously derived for parallel-propagating waves excited by nongyrotropic distributions \cite{sudan1965growing,brinca1993on,cao1995nongyrotropy} to arbitrary propagation angles.

Finally, the pickup process at outer planet moons is further complicated by the coexistence of multiple ion species. These species introduce additional characteristic frequencies to the system and may give rise to new wave modes beyond those excited by each species independently \cite<e.g.,>[]{brinca1989influence}. Extending the present kinetic simulation framework to a multi-species setting would allow one to explore how the nongyrotropic distributions of different ion species interact, compete, or cooperate as free-energy sources for wave growth. The instability criterion derived here [$2 M_{\mathrm{A}}^2 (1 - \eta_{\mathrm{cr}}) / (3 \beta_{\mathrm{cr}}) > 1$] would also need to be generalized to account for the relative densities and mass ratios of the different species. Together, these extensions would build toward a more complete and realistic picture of the ion pickup process at moons such as Io and Europa, where heavy ion species such as O$^+$ and S$^+$ play an important role.

\appendix
\section{Estimate of the ion Bernstein wave dispersion relation}\label{app:ibw}
The most accurate dispersion relation for ion Bernstein waves would require the ion velocity distribution sampled on 3D velocity-space grids from the hybrid-kinetic simulations, which is a project in its own right. Here we instead estimate the dispersion relation using a Maxwellian centered in the plasma rest frame, with a thermal velocity set equal to half of the separation between the corotating ambient and pickup ion populations. This estimate is used in Figure~\ref{fig:disps}(c) of the main text.

Ion Bernstein waves belong to the general class of extraordinary-mode (X-mode) waves propagating quasi-perpendicular to $\mathbf{B}_0 = B_0 \mathbf{e}_z$. The X-mode dispersion relation at $90^\circ$ propagation angle relative to $\mathbf{B}_0$ reads \cite{bookStix62}
\begin{equation}\label{eq:xmode-disp}
    \varepsilon_{xx}\left(\varepsilon_{yy}-N^2\right)=\varepsilon_{xy}\varepsilon_{yx} ,
\end{equation}
where $N \equiv c k_\perp / \omega$ is the refractive index, $\varepsilon_{ij} = \delta_{ij} + \chi_{ij}$ is the dielectric tensor, and $\chi_{ij}$ is the susceptibility tensor. For a Maxwellian species $s$, the susceptibility tensor is
\begin{align}
\chi_{xx,s}&=-\frac{\omega_{ps}^2}{\omega}\sum_{n=-\infty}^{\infty}\frac{n^2\Gn(\lambda_s)}{\lambda_s(\omega-n\omega_{cs})}, \label{eq:chixx}\\
\chi_{xy,s}&=-\chi_{yx,s}=i\,\frac{\omega_{ps}^2}{\omega}\sum_{n=-\infty}^{\infty}\frac{n\left[\Gn(\lambda_s)-\Gnp(\lambda_s)\right]}{\omega-n\omega_{cs}}, \label{eq:chixy}\\
\chi_{yy,s}&=-\frac{\omega_{ps}^2}{\omega}\sum_{n=-\infty}^{\infty}\frac{n^2\Gn(\lambda_s)/\lambda_s+2\lambda_s\left[\Gn(\lambda_s)-\Gnp(\lambda_s)\right]}{\omega-n\omega_{cs}}, \label{eq:chiyy}
\end{align}
where $\lambda_s\equiv k_\perp^2\rho_s^2$ with $\rho_s^2=T_s/(m_s\omega_{cs}^2)$ (i.e., $\rho_s=v_{th,s}/\omega_{cs}$ and $v_{th,s}^2=T_s/m_s$), $\Gn(\lambda)\equiv I_n(\lambda)e^{-\lambda}$, and $\Gnp(\lambda)\equiv e^{-\lambda}I_n'(\lambda)$. Note that $\Gnp$ denotes this combination and is not the total derivative $d\Gn/d\lambda$.

In the isothermal massless-electron limit adopted in our hybrid-kinetic simulations, we take $m_e\to0$ at fixed $T_e$. Then $\rho_e^2\propto m_e\to0$, so $\lambda_e\to0$, while $\omega_{ce}=-|e|B_0/(m_ec)\to-\infty$ and $\wpe^2=4\pi n_ee^2/m_e\to\infty$, with the combination $\wpe^2/\omega_{ce}=-4\pi n_eec/B_0$ remaining finite. Expanding Equations~\eqref{eq:chixx}--\eqref{eq:chiyy} term by term in this limit shows that $\chi_{xx,e}$ and $\chi_{yy,e}$ both vanish at order $m_e$ (every term in the sums carries either an explicit power of $\lambda_e$ or a suppression by $1/\omega_{ce}$), while $\chi_{xy,e}$ survives at leading order, $\chi_{xy,e}\to i\,\wpe^2/(\omega\omega_{ce})$. This is exactly the classic cold-fluid Hall term, confirming that at $k_\parallel=0$ thermal electron effects drop out to leading order in $m_e$.

We normalize length to the ion inertial length $d_i\equiv c/\wpi$ and time to $\omega_{ci}^{-1}$, and define $x\equiv\omega/\omega_{ci}$ and $\hat k\equiv k_\perp d_i=k_\perp c/\wpi$. Two dimensionless parameters then fully determine the problem:
\begin{equation}
\varepsilon\equiv\left(\frac{\omega_{ci}}{\wpi}\right)^2=\left(\frac{v_{\mathrm{A}}}{c}\right)^2,\qquad \alpha\equiv\frac{v_{thi}}{v_{\mathrm{A}}}\quad\Longrightarrow\quad \frac{\rho_i}{d_i}=\alpha,\ \ \lambda_i=\hat k^2\alpha^2, \label{eq:epsalpha}
\end{equation}
where $v_{\mathrm{A}}=B_0/\sqrt{4\pi n_im_i}$ is the Alfv\'en speed and we used $d_i=v_{\mathrm{A}}/\omega_{ci}$. Using $\wpe^2/\omega_{ce}=-\wpi^2/\omega_{ci}$ for a quasi-neutral plasma, so that $\chi_{xy,e}=-i/(\varepsilon x)$, Equation~\eqref{eq:xmode-disp} becomes
\begin{equation}
    \left(1+\chi_{xx,i}(x)\right)\left(1+\chi_{yy,i}(x)-\frac{\hat k^2}{\varepsilon x^2}\right)=G(x)^2, \label{eq:normfinal}
\end{equation}
with
\begin{align}
\chi_{xx,i}(x)&=-\frac{1}{\varepsilon x}\sum_n\frac{n^2\Gn(\lambda_i)}{\lambda_i(x-n)},\\
\chi_{yy,i}(x)&=-\frac{1}{\varepsilon x}\sum_n\frac{n^2\Gn(\lambda_i)/\lambda_i+2\lambda_i\left[\Gn(\lambda_i)-\Gnp(\lambda_i)\right]}{x-n}, \\
G(x)&=\frac{1}{\varepsilon x}\left[\sum_n\frac{n\left[\Gn(\lambda_i)-\Gnp(\lambda_i)\right]}{x-n}-1\right]. \label{eq:Gnorm}
\end{align}
In the hybrid-kinetic regime $\varepsilon\to0$ is a small parameter; neglecting the displacement current, the ``$1$'' terms in Equation~\eqref{eq:normfinal} are dropped. For each $\hat{k}$ we solve Equation~\eqref{eq:normfinal} for $x$ using a root-finding algorithm.

\begin{figure}[htbp]
    \centering
    \includegraphics[width=0.5\linewidth]{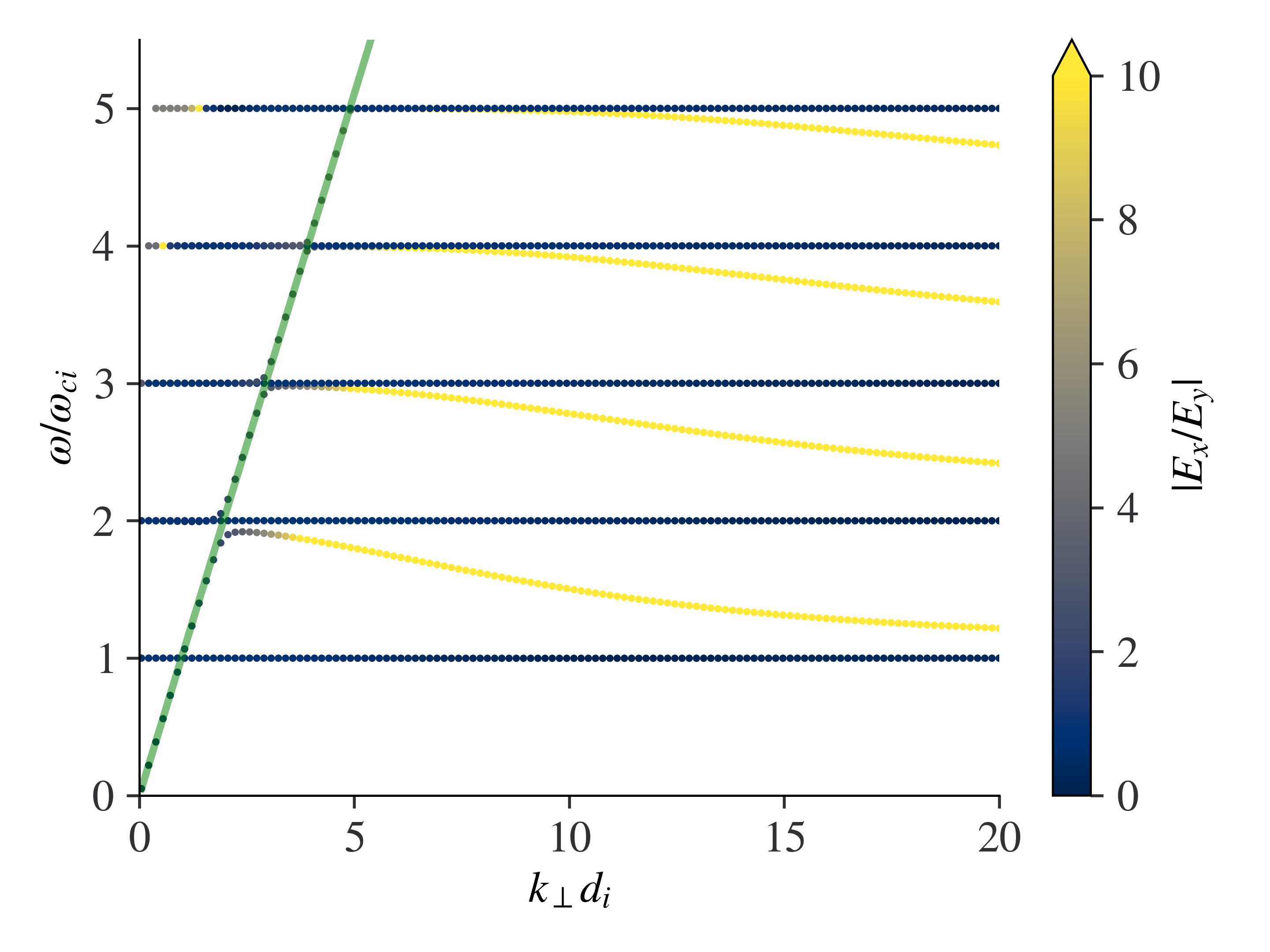}
    \caption{Dispersion relation of the X mode propagating perpendicular to $\mathbf{B}_0$. Each point is color-coded by $\vert E_x / E_y \vert$. The straight green line denotes $\omega / k_\perp = \sqrt{v_{\mathrm{A}}^2 + c_s^2}$.}
    \label{fig:xmode-disp}
\end{figure}

Figure~\ref{fig:xmode-disp} shows the resulting dispersion relation for $\alpha = 0.1125$ (based on half the velocity separation between the corotating ambient and pickup ions) and $\varepsilon = 10^{-4}$, colored by the relative magnitude $\vert E_x / E_y \vert$. Three branches appear: (1) modes near the ion-gyrofrequency harmonics, $\omega / \omega_{ci} \lesssim n$ for $n = 1, 2, \cdots$; (2) a branch aligned with the fast magnetosonic speed, $\omega / k_\perp = \sqrt{v_{\mathrm{A}}^2 + c_s^2}$, where $c_s$ is the ion acoustic speed; and (3) a branch between successive ion-gyrofrequency harmonics, $n < \omega / \omega_{ci} < n+1$. The first two branches are electromagnetic, since $k_\perp \mathbf{e}_x \times \mathbf{E} \neq 0$, while the third is electrostatic, since $k_\perp \mathbf{e}_x \times \mathbf{E} \approx 0$.

In Figure~\ref{fig:disps}(c) of the main text, this dispersion relation is Doppler-shifted to the moon's frame. The branch aligned with the fast magnetosonic speed correctly captures the strongest ion Bernstein wave power at this characteristic propagation speed, whereas the fainter power between ion-gyrofrequency harmonics is not fully reproduced, though the overall trend agrees (i.e., $\omega$ decreases as $k_\perp$ increases). Given that we use an approximate ion velocity distribution, this imperfect agreement between gyrofrequency harmonics is expected.

\section{The field-particle correlation function within the quasi-linear framework\label{sec:fpcorr-ql}}
In the main manuscript, we have defined the field-particle correlation function
\begin{linenomath*}
\begin{align}
\mathcal{C}(\mathbf{v}) = \int \mathrm{d}t \int \mathrm{d}^3 x\, f(\mathbf{x},\mathbf{v},t)\, \mathbf{v}\cdot \mathbf{E}(\mathbf{x},t).
\end{align}
\end{linenomath*}
Here we compute $\mathcal{C}(\mathbf{v})$ in the quasi-linear theoretical framework. Quasi-linear theory decomposes the distribution function into a spatially averaged component $g(\mathbf{v},t)=\langle f(\mathbf{x},\mathbf{v},t)\rangle$ and a rapidly fluctuating component $\delta f(\mathbf{x},\mathbf{v},t)$ associated with waves:
\begin{linenomath*}
\begin{align}
f(\mathbf{x},\mathbf{v},t)
= g(\mathbf{v},t)
+ \int \frac{\mathrm{d}^3 k}{(2\pi)^3}\, e^{i\mathbf{k}\cdot \mathbf{x}}\, \delta f_{\mathbf{k}}(\mathbf{v},t),
\end{align}
\end{linenomath*}
where $\delta f$ has been Fourier-transformed in space. The spatial average is defined as
\begin{linenomath*}
\begin{align}
g(\mathbf{v},t)
= \langle f(\mathbf{x},\mathbf{v},t)\rangle
= \lim_{V \to \infty} \frac{1}{V} \int \mathrm{d}^3 x\, f(\mathbf{x},\mathbf{v},t).
\end{align}
\end{linenomath*}
Given that all wave amplitudes are small, $g(\mathbf{v}, t)$ evolves slowly due to the reaction of the waves back on it.

The electric field is similarly expanded in spatial Fourier modes:
\begin{linenomath*}
\begin{align}
\mathbf{E}(\mathbf{x},t)
= \int \frac{\mathrm{d}^3k}{(2\pi)^3}\, e^{i\mathbf{k}\cdot\mathbf{x}}\, \mathbf{E}_{\mathbf{k}}(t),
\end{align}
\end{linenomath*}
and we take the background (mean) electric field to vanish because we work in the center-of-mass frame.

Since the spatially averaged distribution $g(\mathbf{v},t)$ does not contribute to $\mathcal{C}(\mathbf{v})$, only the perturbed component $\delta f$ enters. Substituting the Fourier representations into the definition of $\mathcal{C}$ gives
\begin{linenomath*}
    \begin{align}\label{eq:corr-func}
        \mathcal{C}(\mathbf{v}) &= \int \mathrm{d}t\, \int \mathrm{d}^3x\, \int \frac{\mathrm{d}^3 k}{(2 \pi)^3} \exp(i \mathbf{k} \cdot \mathbf{x}) \delta f_\mathbf{k} (\mathbf{v}, t) \mathbf{v} \cdot \int \frac{\mathrm{d}^3 k'}{(2 \pi)^3} \exp(i \mathbf{k}' \cdot \mathbf{x}) \mathbf{E}_{\mathbf{k}'}(t) \nonumber \\
        &= \int \mathrm{d}t\, \int \frac{\mathrm{d}^3 k}{(2 \pi)^3} \delta f_\mathbf{k} (\mathbf{v}, t) \mathbf{v} \cdot \int \frac{\mathrm{d}^3 k'}{(2 \pi)^3} \int \mathrm{d}^3x\, \exp[i (\mathbf{k}+\mathbf{k}') \cdot \mathbf{x}] \mathbf{E}_{\mathbf{k}'}(t) \nonumber \\
        &= \int \mathrm{d}t\, \int \frac{\mathrm{d}^3 k}{(2 \pi)^3} \delta f_\mathbf{k} (\mathbf{v}, t) \mathbf{v} \cdot \mathbf{E}_{-\mathbf{k}}(t) \nonumber \\
        &= \int \mathrm{d}t\, \int \frac{\mathrm{d}^3 k}{(2 \pi)^3} \delta f_\mathbf{k} (\mathbf{v}, t) \mathbf{v} \cdot \mathbf{E}_{\mathbf{k}}^*(t) ,
    \end{align}
\end{linenomath*}
where we have used $\int \mathrm{d}^3 x \exp[i (\mathbf{k} + \mathbf{k}') \cdot \mathbf{x}] = (2 \pi)^3 \delta(\mathbf{k} + \mathbf{k}')$ and $\mathbf{E}_{-\mathbf{k}} = \mathbf{E}_\mathbf{k}^*$.

\section{Spatially averaged distribution function\label{sec:avg-gdist}}
Following \citeA{Kennel&Engelmann66}, we Fourier-analyze the Vlasov equation, take the $\mathbf{k}=0$ component, and obtain the evolution equation for $g(\mathbf{v}, t)$:
\begin{linenomath*}
    \begin{align}\label{eq:dtg}
        \partial_t g - \omega_c \partial_\phi g = \lim_{V \to \infty} \frac{1}{V} \int \frac{\mathrm{d}^3 k}{(2 \pi)^3} \hat{P}_{\mathbf{k}} \delta f_{-\mathbf{k}} ,
    \end{align}
\end{linenomath*}
where $\hat{P}_{\mathbf{k}}$ denotes the operator
\begin{linenomath*}
    \begin{align}
        \hat{P}_\mathbf{k} = -\frac{Q}{M} \left(\mathbf{E}_{\mathbf{k}} + \frac{\mathbf{v} \times \mathbf{B}_{\mathbf{k}}}{c}\right) \cdot \partial_{\mathbf{v}} .
    \end{align}
\end{linenomath*}
The term $-\omega_c \partial_\phi g$ represents advection in gyrophase, whereas $\int \mathrm{d}^3k \hat{P}_{\mathbf{k}} \delta f_{-\mathbf{k}}$ represents diffusive scattering in velocity space.

We expand $g(\mathbf{v}, t)$ in gyrophase harmonics, including the advection in gyrophase \cite{sudan1965growing,brinca1993on,cao1995nongyrotropy}:
\begin{linenomath*}
    \begin{align}
        g(\mathbf{v}, t) = \sum_l g_l(v_\perp, v_\parallel, t) e^{i l (\phi + \omega_c t)} .
    \end{align}
\end{linenomath*}
This form of $g(\mathbf{v}, t)$ turns Equation \eqref{eq:dtg} into:
\begin{linenomath*}
    \begin{align}
        \sum_l e^{i l (\phi + \omega_c t)} \partial_t g_l = \lim_{V \to \infty} \frac{1}{V} \int \frac{\mathrm{d}^3 k}{(2 \pi)^3} \hat{P}_{\mathbf{k}} \delta f_{-\mathbf{k}} .
    \end{align}
\end{linenomath*}

\section{Perturbed distribution function\label{sec:pert-df}}
We Fourier-analyze the Vlasov equation, take the $\mathbf{k} \neq 0$ component, and obtain the evolution equation for $\delta f_{\mathbf{k}}$:
\begin{linenomath*}
    \begin{align}\label{eq:df-kennel}
        \hat{L}_{\mathbf{k}} \delta f_\mathbf{k} = \hat{P}_{\mathbf{k}} g ,
    \end{align}
\end{linenomath*}
where $\hat{L}_{\mathbf{k}}$ denotes the operator
\begin{linenomath*}
    \begin{align}
        \hat{L}_{\mathbf{k}} = \partial_t + i \mathbf{k} \cdot \mathbf{v} - \omega_c \partial_\phi .
    \end{align}
\end{linenomath*}
The inverse of $\hat{L}_{\mathbf{k}}$ can be calculated from the orbit integral:
\begin{linenomath*}
    \begin{align}
        \hat{L}_{\mathbf{k}}^{-1} = \int \mathrm{d}^3x \exp(-i \mathbf{k} \cdot \mathbf{x}) \int_{-\infty}^{t} \mathrm{d}t' \int \frac{\mathrm{d}^3k'}{(2\pi)^3} \exp(i \mathbf{k}' \cdot \mathbf{x}' - i \nu_{\mathbf{k}'} t') ,
    \end{align}
\end{linenomath*}
where $\mathbf{x}'$ represents unperturbed particle orbits at time $t'$:
\begin{linenomath*}
    \begin{align}
        & v_\perp' = v_\perp , \label{eq:unpertubed-orbit-vperp} \\
        & \phi' = \phi + \omega_c \tau , \\
        & v_\parallel' = v_\parallel , \\
        & x' = x + \frac{v_\perp}{\omega_c} \sin \phi - \frac{v_\perp}{\omega_c} \sin (\phi + \omega_c \tau), \\
        & y' = y - \frac{v_\perp}{\omega_c} \cos \phi + \frac{v_\perp}{\omega_c} \cos (\phi + \omega_c \tau), \\
        & z' = z - v_\parallel \tau , \label{eq:unpertubed-orbit-z} , \\
        & \tau = t - t' .
    \end{align}
\end{linenomath*}
In the calculation $\delta f_{\mathbf{k}} = \hat{L}_{\mathbf{k}}^{-1} \hat{P}_{\mathbf{k}} g$, $\hat{P}_{\mathbf{k}} g$ must be evaluated along the unperturbed particle orbits: $\mathbf{v}' = \mathbf{v}'(t')$ and $\mathbf{x}' = \mathbf{x}'(t')$:
\begin{linenomath*}
    \begin{align}\label{eq:dfk}
        \delta f_{\mathbf{k}} &= \hat{L}_{\mathbf{k}}^{-1} \hat{P}_{\mathbf{k}} g \nonumber \\
        &= \int \mathrm{d}^3x \exp(-i \mathbf{k} \cdot \mathbf{x}) \int_{-\infty}^{t} \mathrm{d}t' \int \frac{\mathrm{d}^3k'}{(2\pi)^3} \exp(i \mathbf{k}' \cdot \mathbf{x}' - i \nu_{\mathbf{k}'} t') \hat{P}_{\mathbf{k}'} \sum_l g_l(v_\perp, v_\parallel, t) e^{i l (\phi' + \omega_c t')} \nonumber \\
        &= \sum_l \int \mathrm{d}^3x \exp(-i \mathbf{k} \cdot \mathbf{x}) \int_{0}^{\infty} \mathrm{d}\tau \int \frac{\mathrm{d}^3k'}{(2\pi)^3} \exp(i \mathbf{k}' \cdot \mathbf{x}' - i \nu_{\mathbf{k}'} t' + i l \phi' + i l \omega_c t') \hat{P}_{\mathbf{k}'} g_l(v_\perp, v_\parallel, t) ,
    \end{align}
\end{linenomath*}
where the wave frequency is $\nu_{\mathbf{k}}=\omega_{\mathbf{k}}+i\gamma_{\mathbf{k}}$ (with real $\omega_{\mathbf{k}}$ and $\gamma_{\mathbf{k}}$), and the order of $\hat{P}_{\mathbf{k}} e^{i l \phi'}$ is switched to $e^{i l \phi'} \hat{P}_{\mathbf{k}}$ with the understanding that $\partial_\phi$ in $\hat{P}_{\mathbf{k}}$ will be replaced by $i l$. We take the instantaneous value of $g_l(v_\perp, v_\parallel, t)$ assuming its evolution time scale much longer than the gyroperiod.

We calculate the phase factor in Equation \eqref{eq:dfk}:
\begin{linenomath*}
    \begin{align}\label{eq:phase-factor}
        & \exp(i \mathbf{k}' \cdot \mathbf{x}' - i \nu_{\mathbf{k}'} t' + i l \phi' + i l \omega_c t') \nonumber \\
        =& \exp\left[i \mathbf{k}' \cdot \mathbf{x} - i (\nu_{\mathbf{k}'} - l \omega_c) (t - \tau) + i l (\phi + \omega_c \tau) \right] \nonumber \\
        \times & \exp\left[i\frac{k_\perp' v_\perp}{\omega_c} \sin (\phi - \psi) - i\frac{k_\perp' v_\perp}{\omega_c} \sin (\phi - \psi + \omega_c \tau) - i k_\parallel' v_\parallel \tau \right] \nonumber \\
        =& \exp\left[i \mathbf{k}' \cdot \mathbf{x} - i (\nu_{\mathbf{k}'} - l \omega_c) t + i \nu_{\mathbf{k}'} \tau + i l \phi \right] \nonumber \\
        \times & \sum_{m,n} J_m J_n \exp\left[i(m-n)(\phi - \psi) - i n \omega_c \tau - i k_\parallel' v_\parallel \tau\right] \nonumber \\
        =& \exp\left[i \mathbf{k}' \cdot \mathbf{x} - i (\nu_{\mathbf{k}'} - l \omega_c) t + i l \phi \right] \nonumber \\
        \times & \sum_{m,n} J_m J_n \exp\left[i(m-n)(\phi - \psi) + i (\lambda_{\mathbf{k}'} -n \omega_c) \tau\right] ,
    \end{align}
\end{linenomath*}
where $J_n$ is a Bessel function of integer order with argument $k_\perp' v_\perp/\omega_c$, and the Doppler-shifted frequency is $\lambda_{\mathbf{k}'}=\nu_{\mathbf{k}'}-k_\parallel' v_\parallel$. The gyrophase $\phi$ is defined by $v_x = v_\perp\cos\phi$ and $v_y = v_\perp\sin\phi$. $\psi$ is the angle of $\mathbf{k}_\perp$: $k_x = k_\perp\cos\psi$ and $k_y = k_\perp\sin\psi$.

Plugging the phase factor in Equation \eqref{eq:phase-factor} into Equation \eqref{eq:dfk} and using the identify $\int \mathrm{d}^3x \exp(-i \mathbf{k} \cdot \mathbf{x}) \exp(i \mathbf{k}' \cdot \mathbf{x}) = (2 \pi)^3 \delta(\mathbf{k}' - \mathbf{k})$, we obtain
\begin{linenomath*}
    \begin{align}\label{eq:dfk-2}
        \delta f_{\mathbf{k}} &= \hat{L}_{\mathbf{k}}^{-1} \hat{P}_{\mathbf{k}} g \nonumber \\
        &= \sum_l \exp\left[- i (\nu_{\mathbf{k}} - l \omega_c) t + i l \phi \right] \sum_{m,n} J_m J_n e^{i(m-n)(\phi - \psi)} \nonumber \\
        &\times \int_0^\infty \mathrm{d}\tau \exp[i (\lambda_{\mathbf{k}} -n \omega_c) \tau] \hat{P}_{\mathbf{k}} g_l .
    \end{align}
\end{linenomath*}
In the following calculation of $\delta f_{\mathbf{k}}$, the detailed form of the operator $\hat{P}_{\mathbf{k}}$ is used \cite{Kennel&Engelmann66}:
\begin{linenomath*}
    \begin{align}
        \hat{P}_{\mathbf{k}} &= -\frac{Q}{M} e^{i(\phi-\psi)} \left[\frac{E_{\mathbf{k}}^r e^{i\psi}}{\sqrt{2}} \left(\hat{G}_{\mathbf{k}} + \frac{i \lambda_{\mathbf{k}}}{\nu_{\mathbf{k}} v_\perp} \partial_\phi\right) + \frac{k_\perp E_{\mathbf{k}}^\parallel}{2 \nu_{\mathbf{k}}} \left(\hat{H} + i\frac{v_\parallel}{v_{\perp}} \partial_\phi\right)\right] \nonumber \\
        &- \frac{Q}{M} e^{-i(\phi-\psi)} \left[\frac{E_{\mathbf{k}}^l e^{-i\psi}}{\sqrt{2}} \left(\hat{G}_{\mathbf{k}} - \frac{i \lambda_{\mathbf{k}}}{\nu_{\mathbf{k}} v_\perp} \partial_\phi\right) + \frac{k_\perp E_{\mathbf{k}}^\parallel}{2 \nu_{\mathbf{k}}} \left(\hat{H} - i\frac{v_\parallel}{v_{\perp}} \partial_\phi\right)\right] \nonumber \\
        &- \frac{Q}{M} \left[\frac{i k_\perp}{\nu_{\mathbf{k}}} \left(\frac{E_{\mathbf{k}}^l e^{-i\psi} - E_{\mathbf{k}}^r e^{i\psi}}{\sqrt{2}}\right) \partial_\phi + E_{\mathbf{k}}^\parallel \partial_{v_\parallel}\right] ,
    \end{align}
\end{linenomath*}
\begin{linenomath*}
    \begin{align}
        \hat{G}_{\mathbf{k}} = \partial_{v_\perp} - \frac{k_\parallel}{\nu_{\mathbf{k}}} \hat{H} ,
    \end{align}
\end{linenomath*}
\begin{linenomath*}
    \begin{align}
        \hat{H} = v_\parallel \partial_{v_\perp} - v_\perp \partial_{v_\parallel} .
    \end{align}
\end{linenomath*}

Noticing $\phi' = \phi + \omega_c \tau$ along particle orbits in $\hat{P}_{\mathbf{k}}$ and performing the orbit integral parameterized by $\tau$ in Equation \eqref{eq:dfk-2}, we obtain
\begin{linenomath*}
    \begin{align}\label{eq:deltaf-nongyro}
        \delta f_{\mathbf{k}} =& -\frac{i Q}{M} \sum_l \exp\left[- i (\nu_{\mathbf{k}} - l \omega_c) t + i l \phi \right] \sum_{m,n} \frac{J_m e^{i(m-n)(\phi - \psi)}}{\lambda_{\mathbf{k}} - n \omega_c} \nonumber \\
        \cdot& \left[\frac{E_\mathbf{k}^r e^{i \psi} J_{n+1} + E_{\mathbf{k}}^l e^{-i \psi} J_{n-1}}{\sqrt{2}} \hat{G}_\mathbf{k} + E_\mathbf{k}^\parallel J_{n} \left(\partial_{v_\parallel} + \frac{n \omega_c}{\nu_\mathbf{k} v_\perp} \hat{H}\right) \right. \nonumber \\
        &\left. - \frac{E_\mathbf{k}^r e^{i \psi} J_{n+1} - E_{\mathbf{k}}^l e^{-i \psi} J_{n-1}}{\sqrt{2}} \frac{l \lambda_\mathbf{k}}{\nu_\mathbf{k} v_\perp} + E_\mathbf{k}^\parallel J'_{n} \frac{l k_\perp v_\parallel}{\nu_\mathbf{k} v_\perp} +\frac{E_\mathbf{k}^r e^{i \psi} - E_\mathbf{k}^l e^{-i \psi}}{\sqrt{2}} J_{n} \frac{l k_\perp}{\nu_\mathbf{k}} \right] g_l .
    \end{align}
\end{linenomath*}
It is important to note that $\mathbf{E}_{\mathbf{k}}e^{-i \nu_\mathbf{k}t}$ drives $\delta f_{\mathbf{k}}$ at multiple frequencies separated by $\omega_c$: $\delta f_{\mathbf{k}} e^{-i (\nu_\mathbf{k} - l \omega_c) t}$, so that $\mathbf{E}_{\mathbf{k}}(\nu_\mathbf{k})$ naturally couples to $\mathbf{E}_{\mathbf{k}}(\nu_\mathbf{k} - l \omega_c)$ with $l\neq 0$. This result agrees with previous studies \cite{sudan1965growing,brinca1993on}, which focus on the excitation of parallel-propagating waves for nongyrotropic distributions.

For gyrotropic velocity distributions where only the $l = 0$ term contributes, we recover the Kennel-Engelmann result~\cite{Kennel&Engelmann66}:
\begin{linenomath*}
    \begin{align}
        \delta f_\mathbf{k} =& -\frac{i Q}{M} e^{-i \nu_{\mathbf{k}} t} \sum_{n,m} \frac{J_m e^{i(m-n)(\phi - \psi)}}{\lambda_\mathbf{k} - n \omega_c} \nonumber \\
        &\cdot \left[\frac{E_\mathbf{k}^r e^{i \psi} J_{n+1} + E_{\mathbf{k}}^l e^{-i \psi} J_{n-1}}{\sqrt{2}} \hat{G}_\mathbf{k} + E_\mathbf{k}^\parallel J_{n} \left(\partial_{v_\parallel} + \frac{n \omega_c}{\nu_\mathbf{k} v_\perp} \hat{H}\right) \right] g_0 .
    \end{align}
\end{linenomath*}

\section{Calculation of the field-particle correlation function\label{sec:fpcorr-final-formalism}}
Being aware of the coupling between $\mathbf{E}_\mathbf{k}(\nu_\mathbf{k} - l \omega_c)$, we perform the time integral in the field-particle correlation function
\begin{linenomath*}
    \begin{align}\label{eq:corr-func2}
        \mathcal{C}(\mathbf{v}) &= \int \mathrm{d}t\, \int \frac{\mathrm{d}^3 k}{(2 \pi)^3} \delta f_\mathbf{k} (\mathbf{v}, t) \mathbf{v} \cdot \mathbf{E}_{\mathbf{k}}^*(t) \nonumber \\
        &= \int \mathrm{d}t\, \int \frac{\mathrm{d}^3 k}{(2 \pi)^3} \int \mathrm{d}\nu_\mathbf{k} \sum_l \delta f_{\mathbf{k}, l} (\mathbf{v}, \nu_\mathbf{k}) e^{-i (\nu_\mathbf{k} - l \omega_c) t} \mathbf{v} \cdot \int \mathrm{d}\nu_\mathbf{k}' \mathbf{E}_{\mathbf{k}}^*(\nu_\mathbf{k}') e^{i \nu_\mathbf{k}' t} \nonumber \\
        &= 2 \pi \sum_l \int \frac{\mathrm{d}^3 k}{(2 \pi)^3} \int \mathrm{d}\nu_\mathbf{k} \delta f_{\mathbf{k}, l} (\mathbf{v}, \nu_\mathbf{k}) \mathbf{v} \cdot \mathbf{E}_{\mathbf{k}}^*(\nu_\mathbf{k} - l \omega_c) ,
    \end{align}
\end{linenomath*}
where we have used the identity $\int \mathrm{d}t\, e^{-i (\nu_\mathbf{k} - l \omega_c) t} e^{i \nu_\mathbf{k}' t} = 2 \pi \delta(\nu_\mathbf{k}' - \nu_\mathbf{k} + l \omega_c)$.

We calculate the dot product $\mathbf{v} \cdot \mathbf{E}_{\mathbf{k}}^*(\nu_\mathbf{k} - l \omega_c)$ in $\mathcal{C}(\mathbf{v})$:
\begin{linenomath*}
    \begin{align}\label{eq:vdotEstar}
        \mathbf{v} \cdot \mathbf{E}_{\mathbf{k}}^* &= (v_\perp \mathbf{e}_\perp + v_\parallel \mathbf{e}_\parallel) \cdot \left(E_{\mathbf{k}}^l e^{-i \psi} \mathbf{e}_l + E_{\mathbf{k}}^r e^{i \psi} \mathbf{e}_r + E_{\mathbf{k}}^{\parallel} \mathbf{e}_\parallel\right)^* \nonumber \\
        &= (v_\perp \mathbf{e}_\perp + v_\parallel \mathbf{e}_\parallel) \cdot \left((E_{\mathbf{k}}^l e^{-i \psi})^* \mathbf{e}_r + (E_{\mathbf{k}}^r e^{i \psi})^* \mathbf{e}_l + (E_{\mathbf{k}}^{\parallel})^* \mathbf{e}_\parallel\right) \nonumber \\
        &= \frac{v_\perp (E_{\mathbf{k}}^l)^* e^{i(\phi+\psi)} + v_\perp (E_{\mathbf{k}}^r)^* e^{-i(\phi+\psi)}}{\sqrt{2}} + v_\parallel (E_{\mathbf{k}}^{\parallel})^* ,
    \end{align}
\end{linenomath*}
where we have used $\mathbf{e}_l^* = \mathbf{e}_r$, $\mathbf{e}_r^* = \mathbf{e}_l$, $\mathbf{e}_\parallel^* = \mathbf{e}_\parallel$, $\mathbf{e}_l \cdot \mathbf{e}_\perp = e^{-i \phi} / \sqrt{2}$, $\mathbf{e}_r \cdot \mathbf{e}_\perp = e^{i \phi} / \sqrt{2}$, $\mathbf{e}_\parallel \cdot \mathbf{e}_\parallel = 1$, $\mathbf{e}_\perp \cdot \mathbf{e}_\parallel = 0$, and $\mathbf{e}_\parallel \cdot \mathbf{e}_{l, r} = 0$.

Substituting Equations~\eqref{eq:deltaf-nongyro} and \eqref{eq:vdotEstar} into Equation~\eqref{eq:corr-func2} and retaining terms independent of $\phi$, we obtain
\begin{linenomath*}
    \begin{align}
        \mathcal{C}(\mathbf{v}) =& -\frac{2 \pi i Q}{M} \sum_{l,n} \int \frac{\mathrm{d}^3 k}{(2 \pi)^3}  \frac{e^{i l \psi}}{\lambda_\mathbf{k} - n \omega_c} \nonumber \\
        &\cdot \left[ \frac{v_\perp (E_{\mathbf{k}}^l(\nu_\mathbf{k} - l \omega_c) e^{-i \psi})^* J_{n-l-1} e^{i \psi}}{\sqrt{2}} + \frac{v_\perp (E_{\mathbf{k}}^r(\nu_\mathbf{k} - l \omega_c) e^{i \psi})^* J_{n-l+1} e^{-i \psi}}{\sqrt{2}} \right. \nonumber \\
        &\left. + v_\parallel (E_{\mathbf{k}}^{\parallel}(\nu_\mathbf{k} - l \omega_c))^* J_{n-l} \right] \nonumber \\
        &\cdot \left[\frac{E_\mathbf{k}^r(\nu_\mathbf{k}) e^{i \psi} J_{n+1} + E_{\mathbf{k}}^l(\nu_\mathbf{k}) e^{-i \psi} J_{n-1}}{\sqrt{2}} \hat{G}_\mathbf{k} + E_\mathbf{k}^\parallel(\nu_\mathbf{k}) J_{n} \left(\partial_{v_\parallel} + \frac{n \omega_c}{\nu_\mathbf{k} v_\perp} \hat{H}\right) \right. \nonumber \\
        &\left. - \frac{E_\mathbf{k}^r(\nu_\mathbf{k}) e^{i \psi} J_{n+1} - E_{\mathbf{k}}^l(\nu_\mathbf{k}) e^{-i \psi} J_{n-1}}{\sqrt{2}} \frac{l \lambda_\mathbf{k}}{\nu_\mathbf{k} v_\perp} + E_\mathbf{k}^\parallel(\nu_\mathbf{k}) J'_{n} \frac{l k_\perp v_\parallel}{\nu_\mathbf{k} v_\perp} \right. \nonumber \\
        &\left.+\frac{E_\mathbf{k}^r(\nu_\mathbf{k}) e^{i \psi} - E_\mathbf{k}^l(\nu_\mathbf{k}) e^{-i \psi}}{\sqrt{2}} J_{n} \frac{l k_\perp}{\nu_\mathbf{k}}\right] g_l \nonumber \\
        \equiv& \sum_{l,n} \sum_{\sigma, \sigma'} \int \frac{\mathrm{d}^3 k}{(2 \pi)^3} \mathcal{K}_{\sigma \sigma'}^{(n, l)} E_{\mathbf{k}}^\sigma (\nu_{\mathbf{k}}) E_{\mathbf{k}}^{\sigma' *} (\nu_{\mathbf{k}} - l \omega_c),
    \end{align}
\end{linenomath*}
where $\sigma, \sigma' \in \{l, r, \parallel\}$, and $\mathcal{K}_{\sigma \sigma'}^{(n, l)}$ is the coupling coefficient of resonance order $n$ and gyrophase harmonic $l$ that couples the field polarization $\sigma$ at frequency $\nu_{\mathbf{k}}$ to polarization $\sigma'$ at frequency $\nu_{\mathbf{k}} - l\omega_c$. The explicit expressions for each component are:
\begin{linenomath*}
    \begin{align}
        \mathcal{K}_{ll}^{(n,l)} &= \mathcal{A}_{n,l} \frac{v_\perp e^{i\psi} J_{n-l-1}}{2} G_l , \\
        \mathcal{K}_{lr}^{(n,l)} &= \mathcal{A}_{n,l} \frac{v_\perp e^{-i 3 \psi} J_{n-l+1}}{2} G_l , \\
        \mathcal{K}_{l\parallel}^{(n,l)} &= \mathcal{A}_{n,l} \frac{v_\parallel e^{-i \psi} J_{n-l}}{\sqrt{2}} G_l , \\
        \mathcal{K}_{rl}^{(n,l)} &= \mathcal{A}_{n,l} \frac{v_\perp e^{i 3 \psi} J_{n-l-1}}{2} G_r , \\
        \mathcal{K}_{rr}^{(n,l)} &= \mathcal{A}_{n,l} \frac{v_\perp e^{-i \psi} J_{n-l+1}}{2} G_r , \\
        \mathcal{K}_{r\parallel}^{(n,l)} &= \mathcal{A}_{n,l} \frac{v_\parallel e^{i \psi} J_{n-l}}{\sqrt{2}} G_r , \\
        \mathcal{K}_{\parallel l}^{(n,l)} &= \mathcal{A}_{n,l} \frac{v_\perp e^{i 2\psi} J_{n-l-1}}{\sqrt{2}} G_\parallel , \\
        \mathcal{K}_{\parallel r}^{(n,l)} &= \mathcal{A}_{n,l} \frac{v_\perp e^{-i 2\psi} J_{n-l+1}}{\sqrt{2}} G_\parallel , \\
        \mathcal{K}_{\parallel \parallel}^{(n,l)} &= \mathcal{A}_{n,l} v_\parallel J_{n-l} G_\parallel ,
    \end{align}
\end{linenomath*}
where
\begin{linenomath*}
    \begin{align}
        \mathcal{A}_{n,l} &= -\frac{2 \pi i Q}{M} \frac{e^{i l \psi}}{\lambda_\mathbf{k} - n \omega_c} , \\
        G_l &= \left( J_{n-1} \hat{G}_{\mathbf{k}} + J_{n-1} \frac{l \lambda_\mathbf{k}}{\nu_\mathbf{k} v_\perp} -J_n \frac{l k_\perp}{\nu_\mathbf{k}} \right) g_l , \\
        G_r &= \left( J_{n+1} \hat{G}_{\mathbf{k}} - J_{n+1} \frac{l \lambda_\mathbf{k}}{\nu_\mathbf{k} v_\perp} +J_n \frac{l k_\perp}{\nu_\mathbf{k}} \right) g_l , \\
        G_\parallel &= \left[J_{n} \left(\partial_{v_\parallel} + \frac{n \omega_c}{\nu_\mathbf{k} v_\perp} \hat{H}\right) + J'_{n} \frac{l k_\perp v_\parallel}{\nu_\mathbf{k} v_\perp} \right] g_l .
    \end{align}
\end{linenomath*}

\section{Field-particle correlation coefficients for ion-Bernstein and mirror-mode waves}\label{sec:fpcorr-ibw-mmw}
\begin{figure}[tphb]
    \centering
    \includegraphics[width=\linewidth]{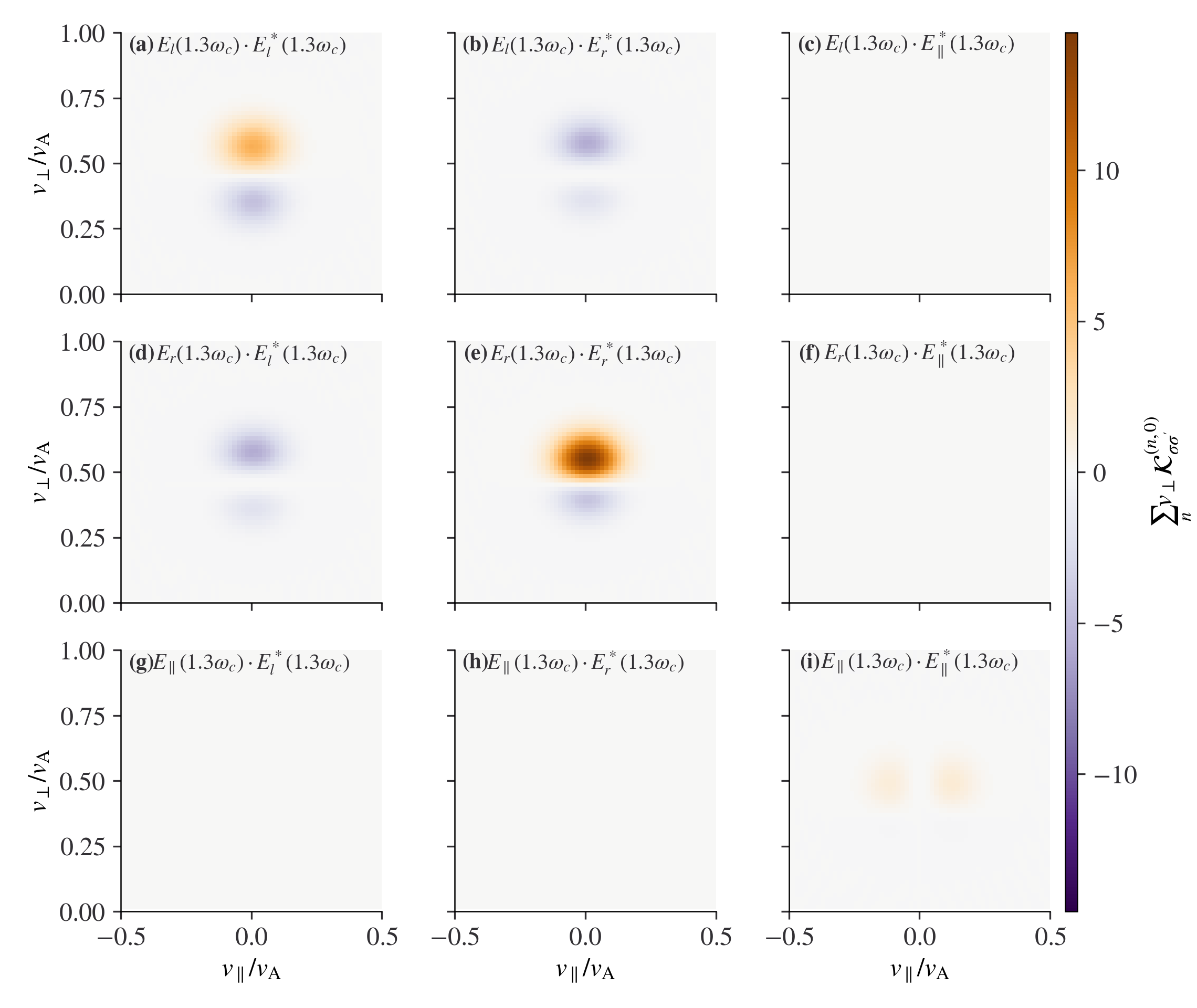}
    \caption{Coupling coefficients of the field-particle correlation function for the fastest growing ion Bernstein mode with gyrophase harmonic distribution $g_{0}$. Panel layout and notation are the same as in Figure \ref{fig:fpcorr-ibw-lminus1}.}
    \label{fig:fpcorr-ibw-l0}
\end{figure}

\begin{figure}[tphb]
    \centering
    \includegraphics[width=\linewidth]{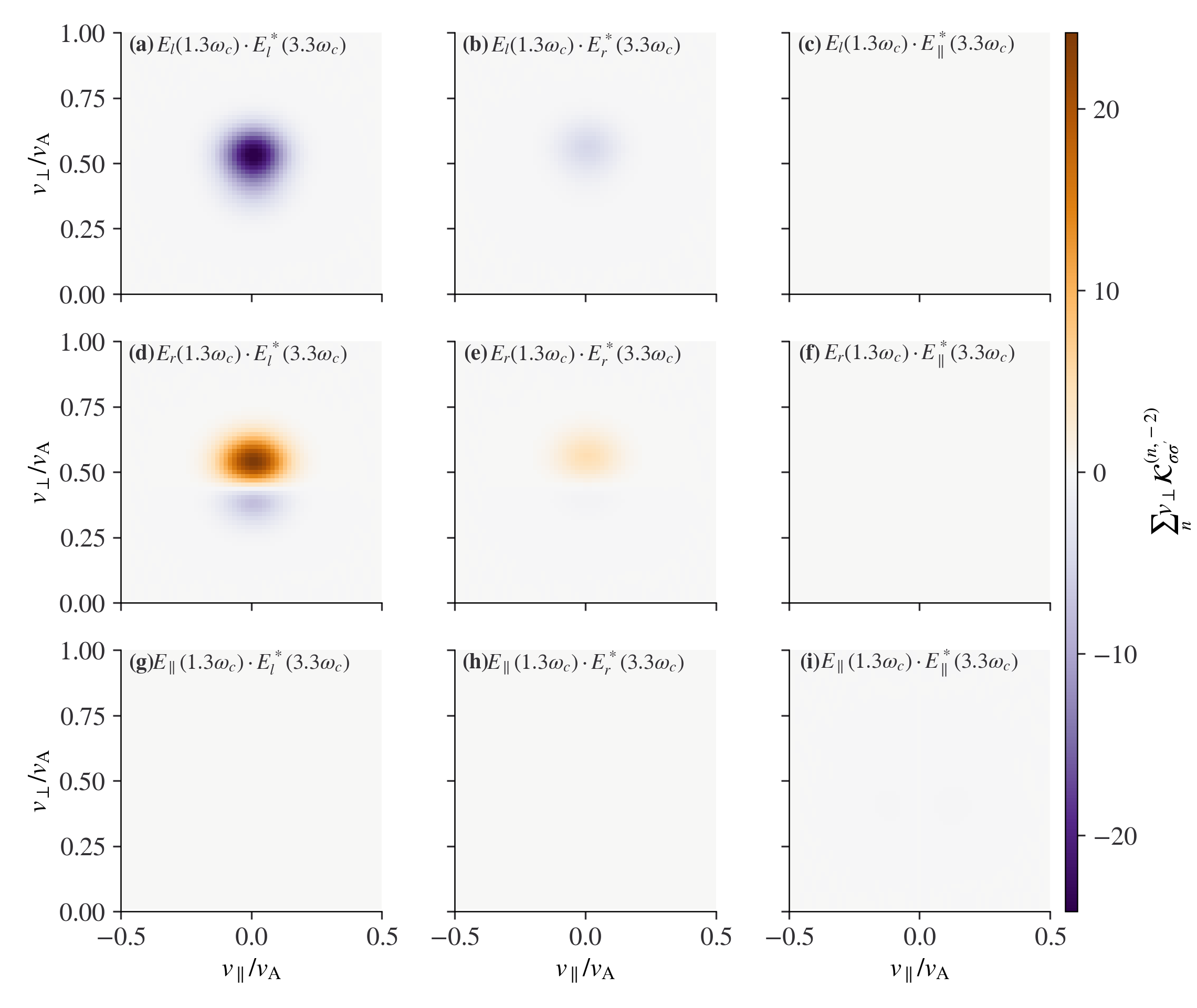}
    \caption{Coupling coefficients of the field-particle correlation function for the fastest growing ion Bernstein mode with gyrophase harmonic distribution $g_{-2}$. Panel layout and notation are the same as in Figure \ref{fig:fpcorr-ibw-lminus1}.}
    \label{fig:fpcorr-ibw-lminus2}
\end{figure}

\begin{figure}[tphb]
    \centering
    \includegraphics[width=\linewidth]{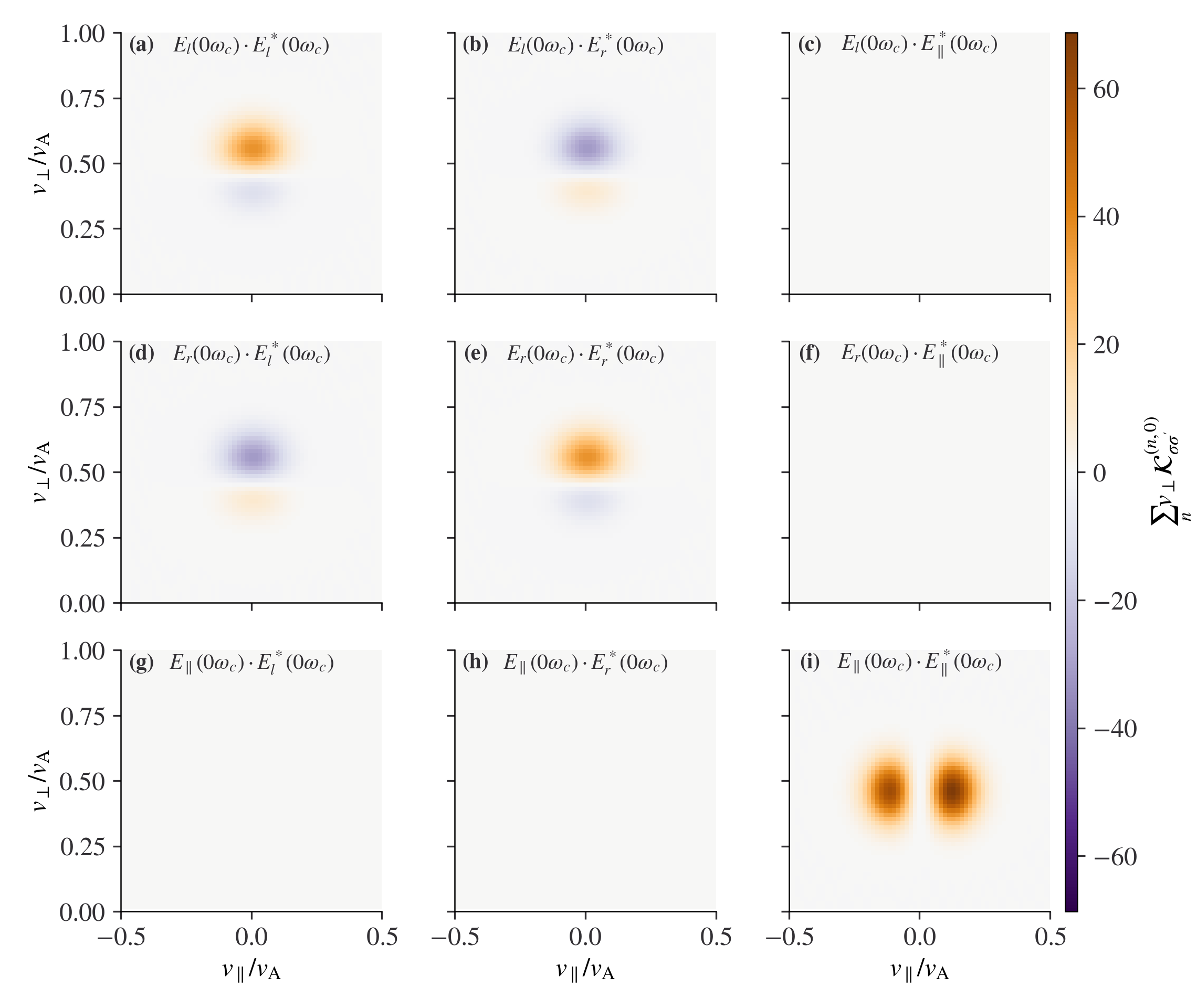}
    \caption{Coupling coefficients of the field-particle correlation function for the fastest growing mirror mode with gyrophase harmonic distribution $g_{0}$. Panel layout and notation are the same as in Figure \ref{fig:fpcorr-ibw-lminus1}.}
    \label{fig:fpcorr-mmw-l0}
\end{figure}

\begin{figure}[tphb]
    \centering
    \includegraphics[width=\linewidth]{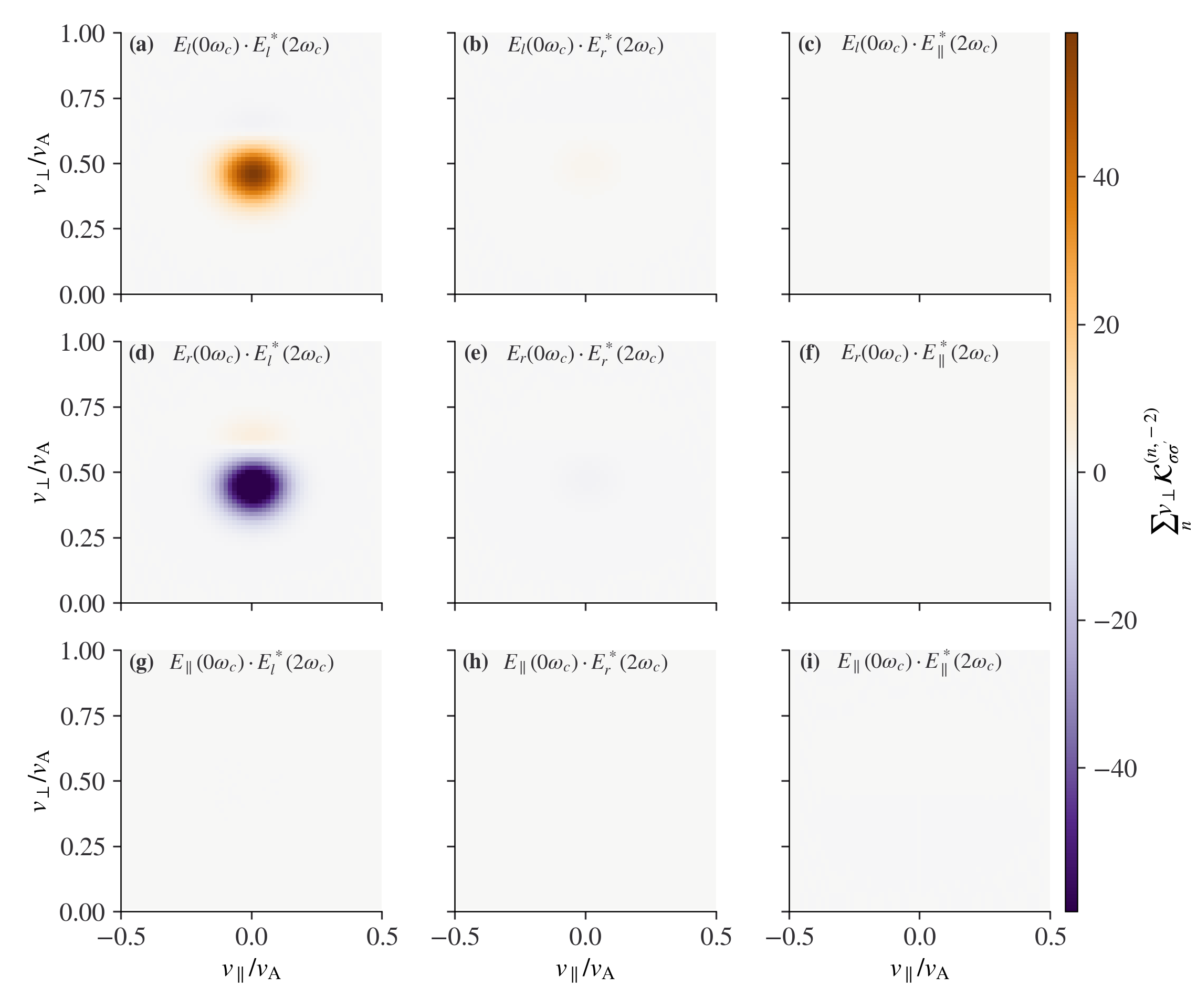}
    \caption{Coupling coefficients of the field-particle correlation function for the fastest growing mirror mode with gyrophase harmonic distribution $g_{-2}$. Panel layout and notation are the same as in Figure \ref{fig:fpcorr-ibw-lminus1}.}
    \label{fig:fpcorr-mmw-lminus2}
\end{figure}

This section presents coupling coefficients of the field-particle correlation between additional wave modes and gyrophase harmonic distributions, complementing Figures~\ref{fig:fpcorr-ibw-lminus1} and \ref{fig:fpcorr-mmw-lminus1} in the main text. Figures~\ref{fig:fpcorr-ibw-l0} and \ref{fig:fpcorr-ibw-lminus2} show results for the fastest growing ion Bernstein mode with gyrophase harmonic distributions $g_0$ and $g_{-2}$, respectively, while Figures~\ref{fig:fpcorr-mmw-l0} and \ref{fig:fpcorr-mmw-lminus2} show the corresponding results for the fastest growing mirror mode. Of particular note, the gyrotropic distribution $g_0$ exhibits persistent Landau damping for both wave modes [Figures~\ref{fig:fpcorr-ibw-l0}(i) and \ref{fig:fpcorr-mmw-l0}(i)], which may account for the slow decay of the compressional magnetic field $B_\parallel$ observed after wave saturation [Figure~\ref{fig:energy-budget}(b)].

\section*{Open Research Section}
The simulation data and Jupyter notebooks used in the study are available at Dryad via \citeA{An2026ion}.

\section*{Conflict of Interest declaration}
The authors declare there are no conflicts of interest for this manuscript.

\acknowledgments
This work was supported by NASA Juno mission and THEMIS/ARTEMIS grant NAS5-02099. We would like to acknowledge high-performance computing support from Derecho (\url{https://doi.org/10.5065/qx9a-pg09}) provided by NCAR's Computational and Information Systems Laboratory, sponsored by the National Science Foundation \cite{derecho}. Xin An and Anton Artemyev acknowledge support from the International Space Science Institute (ISSI) in Bern through the international team project ``Beyond Diffusion: Advancing Earth's Radiation Belt Models with Nonlinear Dynamics'' (ISSI Team Project \#25-640). Miranda Chang acknowledges the support by the National Science Foundation Graduate Research Fellowship Program under Grant No.~DGE-2444110. Any opinions, findings, and conclusions or recommendations expressed in this material are those of the authors and do not necessarily reflect the views of the National Science Foundation.

%
%


%
%
%
%
%

\end{document}